\documentclass[showpacs,showkeys,notitlepage,superscriptaddress,preprintnumbers,secnumarabic,nofootinbib,nobibnotes,onecolumn]{revtex4-1}
\usepackage{amsmath,amssymb,bm,mathrsfs}
\usepackage{srcltx,hyperref}
\usepackage{epsfig}
\usepackage{slashed}
\usepackage{bbold}
\usepackage{psfrag}
\usepackage{color}
\PassOptionsToPackage{caption=false}{subfig}
\usepackage{subfig}
\usepackage{multirow}
\usepackage{booktabs}
\usepackage{hyperref}

\newcommand{\ccbar}{{c\bar{c}}}
\newcommand{\bbbar}{{b\bar{b}}}
\newcommand{\BR}{{\rm BR}}

\newcommand{\cO}{\mathcal{O}}
\newcommand{\cL}{\mathcal{L}}

\newcommand{\cR}{\mathcal{R}}
\newcommand{\abs}[1]{\left\lvert#1\right\rvert}

\newcommand{\beq}{\begin{equation}}
\newcommand{\eeq}{\end{equation}}

\newcommand{\fb}{{\rm fb}}

\begin{document}

\preprint{KEK-TH-1842}

\title{Prospects for measuring the Higgs coupling to light quarks} 

\author{Gilad Perez}
\email{gilad.perez@weizmann.ac.il}
\affiliation{Department of Particle Physics and Astrophysics,
Weizmann Institute of Science, Rehovot 7610001, Israel} 
\author{Yotam Soreq}
\email{yotam.soreq@weizmann.ac.il}
\affiliation{Department of Particle Physics and Astrophysics,
Weizmann Institute of Science, Rehovot 7610001, Israel} 
\author{Emmanuel Stamou}
\email{emmanuel.stamou@weizmann.ac.il}
\affiliation{Department of Particle Physics and Astrophysics,
Weizmann Institute of Science, Rehovot 7610001, Israel} 
\author{Kohsaku Tobioka}
\email{kohsakut@post.tau.ac.il}
\affiliation{Department of Particle Physics and Astrophysics,
Weizmann Institute of Science, Rehovot 7610001, Israel} 
\affiliation{Theory Center, High Energy Accelerator Research Organization (KEK), Tsukuba 305-0801, Japan}
\affiliation{Raymond and Beverly Sackler School of Physics and Astronomy, \\
Tel-Aviv University, Tel-Aviv 6997801, Israel}

\begin{abstract}
\vspace*{1cm}
\centerline{\bf Abstract}
\vspace*{.1cm}
We discuss the prospects to probe the light-quark Yukawa couplings to the Higgs boson. 
The Higgs coupling to the charm quark can be probed both via inclusive and exclusive approaches. 
On the inclusive frontier, we use our recently proposed method together with published 
experimental studies for the sensitivity of the Higgs coupling to bottom quarks to find that the high-luminosity LHC can 
be sensitive to modifications of the charm Yukawa of the order of a few times its standard model~(SM) value.
We also present a preliminary study of this mode for a $100$\,TeV hadronic machine 
(with similar luminosity) and find that the bound can be further improved, possibly 
within the reach of the expected signal in the SM. 
On the exclusive frontier, we use the recent ATLAS search for charmonia and photon 
final state. This study yields the first measurement of the background relevant to these modes. 
Using this background measurement we project that at the high-luminosity LHC, 
unless the analysis strategy is changed, the sensitivity of the exclusive final 
state to the charm Yukawa to the charm Yukawa will be rather poor, of the order 
of $50$ times the SM coupling. 
We then use a Monte-Carlo study to rescale the above backgrounds to the $h\to \phi \gamma$ 
case and obtain a much weaker sensitivity to the strange Yukawa, of order of $3000$ 
times the SM value.  
We briefly speculate what would be required to improve the prospects of the exclusive modes. 
\end{abstract}
\maketitle
%\nopagebreak
% \pacs{}
% \keywords{} 

%%%%%%%%%%%%%%%%%%%%%%%%%%%%%%%%%%%%%%%%%%%%%%%%
\section{Introduction}
%%%%%%%%%%%%%%%%%%%%%%%%%%%%%%%%%%%%%%%%%%%%%%%%

Now that the Higgs particle has been discovered~\cite{Aad:2012tfa,Chatrchyan:2012ufa} the standard 
model~(SM) is complete.
It has a minimal scalar sector of electroweak~(EW) symmetry breaking and is a theory consistent
up to very high scales. 
Furthermore, the SM enjoys a set of accidental (exact and approximate) symmetries leading to: 
baryon$-$lepton number conservation, suppression of processes involving flavor-changing 
neutral currents and CP~violation.
Nevertheless, the flavor sector of the SM has a very particular structure.
The Higgs couplings depend linearly on the masses, which implies that most of the Yukawas are
small and hierarchical, leading to the SM flavor puzzle. 
However, at present there is no strong direct evidence for the validity of this particular
structure. 
For instance, it is not impossible that the masses of the first two generation fermions
originate from a different source of EW symmetry breaking
thus leading to deviations from the simple SM relation between fermion masses and 
their coupling to the Higgs.

With new physics it is actually easy to obtain enhancements or suppressions in the strengths 
of Higgs to light-quark interactions. 
Furthermore, as the Higgs is rather light, within the SM it can only decay to particles that 
interact very weakly with it, with the dominant decay to $b \bar b$. 
A deformation of the Higgs couplings to the lighter SM particles, 
say for the charm or other quarks (see Refs.~\cite{Delaunay:2013iia,Delaunay:2013pwa,Blanke:2013uia,Mahbubani:2012qq,Kagan:2009bn,Dery:2013aba,Giudice:2008uua,DaRold:2012sz,Dery:2014kxa,Bishara:2015cha}), 
could compete with the Higgs--bottom coupling and would lead to a dramatic change of the 
Higgs phenomenology \cite{Delaunay:2013pja}.

Our knowledge of the Higgs Yukawa couplings is mainly on the third-generation 
charged fermions. 
Though not yet fully conclusive, it is consistent with the SM Higgs mechanism of 
fermion-mass generation~\cite{ATLAStth, Aad:2014xzb, Aad:2015vsa, Khachatryan:2014qaa, 
Chatrchyan:2013zna, Chatrchyan:2014nva}. 
Regarding the first two generations, at present, we only have a rather weak 
upper bound on the corresponding signal strengths of muons and electrons~\cite{Aad:2014xva,Khachatryan:2014aep} 
\begin{equation} \label{eq:muemu}
	\mu_{\mu}\leq 7 \, ,  \quad\quad
	\mu_{e}\leq 4 \times 10^5 \,,
\end{equation}
at 95\%~Confidence Level (CL) where $\mu_{f}\equiv \frac{\sigma_h}{\sigma_h^{\rm SM}}\,
\frac{\BR_{f\bar f}}{\BR_{ f\bar f}^{\rm SM}}$ with $\sigma_h$ standing for the Higgs-production 
cross section, $\BR_{X}\equiv\BR(h\to X)$ and the SM script indicating the SM case.
In addition, in Ref.~\cite{Perez:2015aoa} we recasted the ATLAS~\cite{Aad:2014xzb} and 
CMS~\cite{Chatrchyan:2013zna} studies of $Vh (b\bar b)$ and obtained a first direct bound on 
the charm signal strength,
\begin{equation} \label{eq:muemu}
	\mu_{c}\leq 270\,,
\end{equation}
at $95$\%\,CL. These bounds are very weak, yet they are sufficient to exclude Higgs-coupling universality 
to quarks and charged leptons.

Let us summarise the current status of the theoretical and experimental activity 
relevant to probing Higgs to light-quark couplings.
On the theoretical frontier, it was demonstrated in Refs.~\cite{Delaunay:2013pja,Perez:2015aoa} that 
inclusive charm-tagging enables the LHC experiments to constrain the charm Yukawa coupling.
Furthermore, it was shown that the Higgs--charm coupling may be probed by looking at 
exclusive decay modes involving a $c$-$\bar c$ vector meson and a photon~\cite{Bodwin:2013gca}. 
This makes the charm Yukawa coupling rather special among the light quarks as it can be probed 
both with inclusive and exclusive approaches. 
A similar mechanism, based on exclusive decays to light-quark states and gauge bosons $\gamma/W/Z$, 
was shown to yield a potential access to the Higgs--light-quark couplings~\cite{Kagan:2014ila}. 
(See also Refs.~\cite{Isidori:2013cla,Mangano:2014xta,Huang:2014cxa, Grossmann:2015lea} for studies 
of exclusive EW gauge-boson decays and new-physics searches.) 
On the experimental side, ATLAS recently published two searches for supersymmetry
\cite{Aad:2014nra, Aad:2015gna}, which employ charm-tagging ($c$-tagging) \cite{ATL-PHYS-PUB-2015-001}.  
ATLAS further published an analysis that focuses on Higgs decays to quarkonia
({\it e.g.} $J/\psi$, $\Upsilon$) plus a photon final
states~\cite{Aad:2015sda}.\footnote{Note added: during the reviewing process of this article
the CMS collaboration presented results on a similar analysis \cite{Khachatryan:2015lga}.}

In this paper, we discuss the prospects to probe light-quark Yukawa couplings 
to the Higgs boson at LHC run II, the high-luminosity LHC~(HL-LHC) and possible 
future colliders.
We will demonstrate that the HL-LHC can reach a sensitivity for the charm 
Yukawa up to few times the SM value by using the inclusive method recently proposed by 
us~\cite{Perez:2015aoa}.
We also present a preliminary study of this mode for a $100$\,TeV hadronic machine assuming 
similar luminosity as at the HL-LHC. 
We shall find that the sensitivity can be further improved, possibly probing the
signal expected in the SM.
On the exclusive frontier we shall use the recent ATLAS result on the search for 
charmonia plus a photon final state \cite{Aad:2015sda}.  
This study yields the first measurement of the background relevant to these modes. 
It is dominated by a jet converted to photon or a real photon plus charmonia production. 
Given this background measurement, we project that, unless the analysis 
strategy is changed, the sensitivity of the exclusive final state
to the charm Yukawa at the HL-LHC will be rather poor, of the order of $50$ times 
the SM coupling. 
We then use a {\tt PYTHIA} simulation to rescale the above backgrounds to the 
$h\to \phi \gamma$ case and obtain a much weaker sensitivity to the strange Yukawa 
of order of $3000$ times the SM value.  

In the following section we discuss the prospects for probing the charm 
Yukawa coupling using charm tagging. 
In Section~\ref{sec:Exclusive} we focus on the exclusive case. 
For completeness, we also briefly describe in Section~\ref{sec:ee} the corresponding 
status for $e^+ e^-$ machines. 
We conclude in Section~\ref{sec:conc}.

%%%%%%%%%%%%%%%%%%%%%%%%%%%%%%%%%%%%%%%%%%%%%%%%
\section{Inclusive Analysis} \label{sec:Inclusive}
%%%%%%%%%%%%%%%%%%%%%%%%%%%%%%%%%%%%%%%%%%%%%%%%

We begin by estimating the future sensitivity of the LHC and a future $100$\,TeV $pp$ collider
to probe the $h\to\ccbar$ signal strength, $\mu_c$, and the charm Yukawa, 
$\kappa_c\equiv y_c/y^{\rm SM}_c$, via the inclusive method proposed in
Ref.~\cite{Perez:2015aoa}.
The method takes advantage of the fact that the signal strength in searches
for $h\to b\bar b$ requires two $b$-tagged jets.
In this way, the same analyses are also sensitive to $h\to\ccbar$ events 
because charm jets ($c$-jets) pass the tag criteria with a non-negligible rate.
To account for such events, the $h\to b\bar b$ signal 
strength, $\mu_b$, is extended 
\begin{align} 
	\mu_b \equiv \frac{\sigma_h  \BR_{\bbbar} }{\sigma^{\rm SM}_h \BR_{\bbbar}^{\rm SM}} 
	\, \to  \, 
	\frac{\sigma_h  \BR_{\bbbar} \epsilon_{b_1}\epsilon_{b_2} + \sigma_h  \BR_{\ccbar} \epsilon_{c_1}\epsilon_{c_2}   }
	{\sigma^{\rm SM}_h \BR_{\bbbar}^{\rm SM}\epsilon_{b_1}\epsilon_{b_2}+\sigma^{\rm SM}_h \BR_{\ccbar}^{\rm SM}\epsilon_{c_1}\epsilon_{c_2}}
=	\left( \mu_b + \frac{\BR_{\ccbar}^{\rm SM}}{\BR_{\bbbar}^{\rm SM}} \frac{\epsilon_{c_1}\epsilon_{c_2}}{\epsilon_{b_1}\epsilon_{b_2}} \mu_c\right)
	\Bigg/ 
	\left( 1 + \frac{\BR_{\ccbar}^{\rm SM}}{\BR_{\bbbar}^{\rm SM}} \frac{\epsilon_{c_1}\epsilon_{c_2}}{\epsilon_{b_1}\epsilon_{b_2}} \right)\, . 
	\label{eq:mub2mubmuc}
\end{align}
Here, $\epsilon_{b}$ and $\epsilon_{c}$ are the efficiencies to tag jets 
originating from bottom and charm quarks, respectively. The subscripts~$1$ and~$2$ refer
to the efficiency of tagging the first and second jet, respectively and
$\BR_{\ccbar}^{\rm SM}/\BR_{\bbbar}^{\rm SM}\simeq5\%$~\cite{Heinemeyer:2013tqa}. 
For brevity, we define the ratio of tagging efficiencies, 
$\epsilon_{c/b}^2\equiv (\epsilon_{c_1}\epsilon_{c_2})/(\epsilon_{b_1}\epsilon_{b_2})$.  

An analysis employing a single jet-tagger, such as medium $b$-tagging, constrains  only 
a linear combination of $\mu_{b}$ and $\mu_{c}\,$. 
To be able to separately obtain $\mu_{c}$ we need at least two analyses with 
different ratios, $\epsilon_{c/b}^2$, in order to break the degeneracy in 
Eq.~\eqref{eq:mub2mubmuc}.
The sensitivity is best when in the employed tagger $\epsilon_{c/b}^2$ is large
meaning that many $c$-jets are being tagged.
This situation is realized if we combine $b$-tagging with {\it charm tagging} ($c$-tagging),
which is a jet-tagger optimized for $c$-jets. 
For the $b$-tagger we will always consider the medium $b$-tagging 
working point as in Ref.~\cite{ATL-PHYS-PUB-2014-011}. 
For $c$-tagging we shall use what ATLAS already employed at run~I~\cite{Aad:2014nra, Aad:2015gna,ATL-PHYS-PUB-2015-001}
and refer to it as $c$-tagging~I. Moreover, given the recent installation of the new Insertable B-Layer~(IBL) 
subdetector~\cite{Capeans:1291633} in the ATLAS detector, the capability for 
$c$-tagging is expected to be much improved. Related to this, Ref.~\cite{ATL-PHYS-PUB-2013-010} shows possible  improvement of the life-time resolution  of $B^0_s\to J/\psi\,\phi$ decay by 30\%, thanks to the IBL. 
Thus, we consider two additional $c$-taggings,  
referred to as $c$-tagging~II and $c$-tagging~III. 
All tagging efficiencies are summarized in Table~\ref{tab:efficiencies} in which 
$\epsilon_l$ denotes the efficiency to tag a light jet. 

\begin{table}[t]
\begin{tabular}[t]{l@{\hskip 1.5em}c@{\hskip 1.5em}c@{\hskip 1.5em}c}
			& $\boldsymbol{\epsilon_b}$	
			& $\boldsymbol{\epsilon_c}$	
			& $\boldsymbol{\epsilon_l}$  \\\hline\hline
{\bf $b$-tagging}	& $70\%$	& $20\%$	& $1.25\%$	\\
{\bf $c$-tagging~I}	& $13\%$	& $19\%$	& $0.5\%$  	\\
{\bf $c$-tagging~II}	& $20\%$	& $30\%$	& $0.5\%$	\\
{\bf $c$-tagging~III}	& $20\%$	& $50\%$	& $0.5\%$	\\\hline
\end{tabular}
\caption{The tagging efficiencies for the four jet-taggers used in our analysis.
\label{tab:efficiencies}}
\end{table}

\begin{figure}[t!]
\centering
%\vspace{-10pt}
\includegraphics[width=0.35\textwidth]{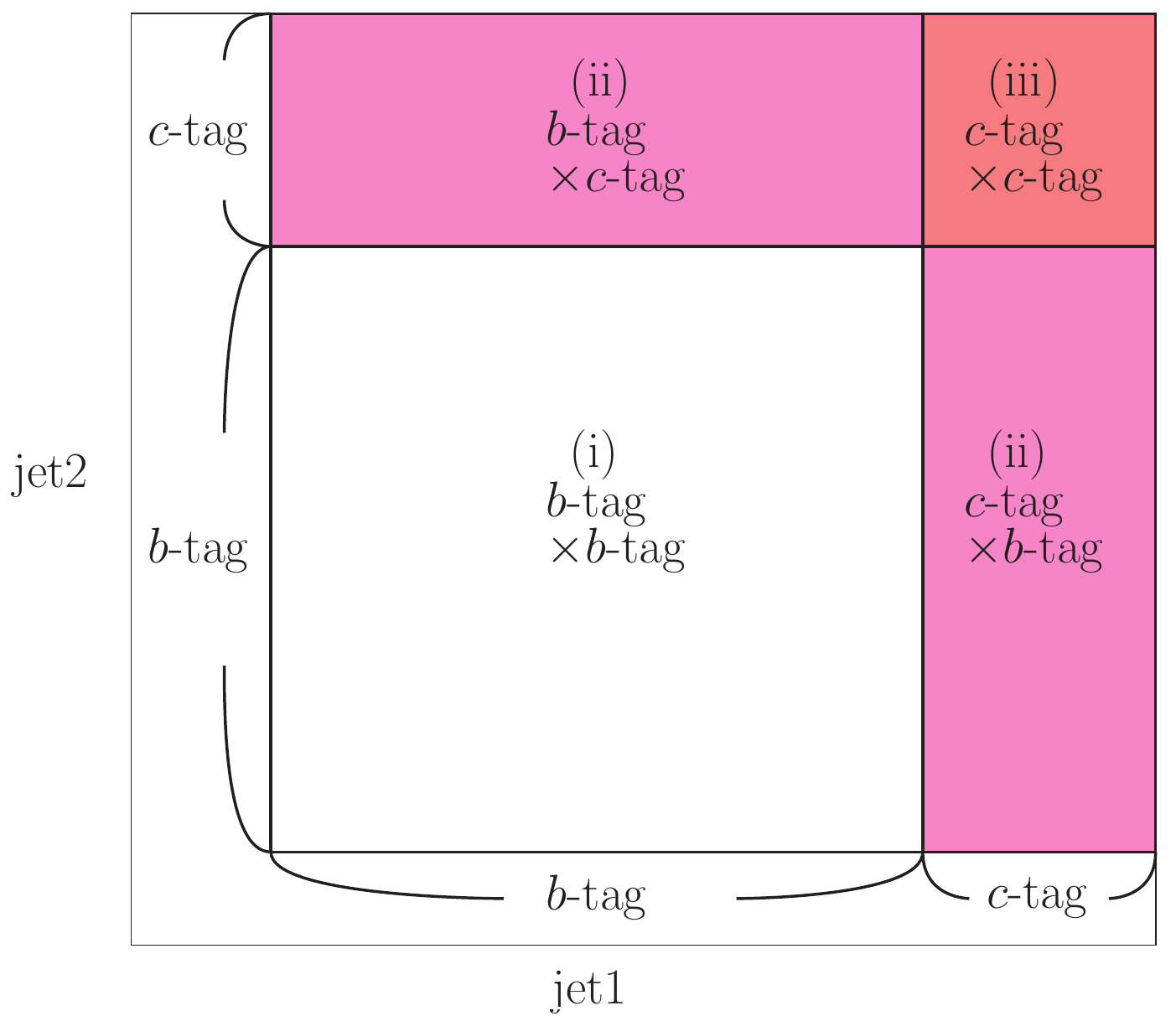}\hspace{2cm}
\includegraphics[width=0.35\textwidth]{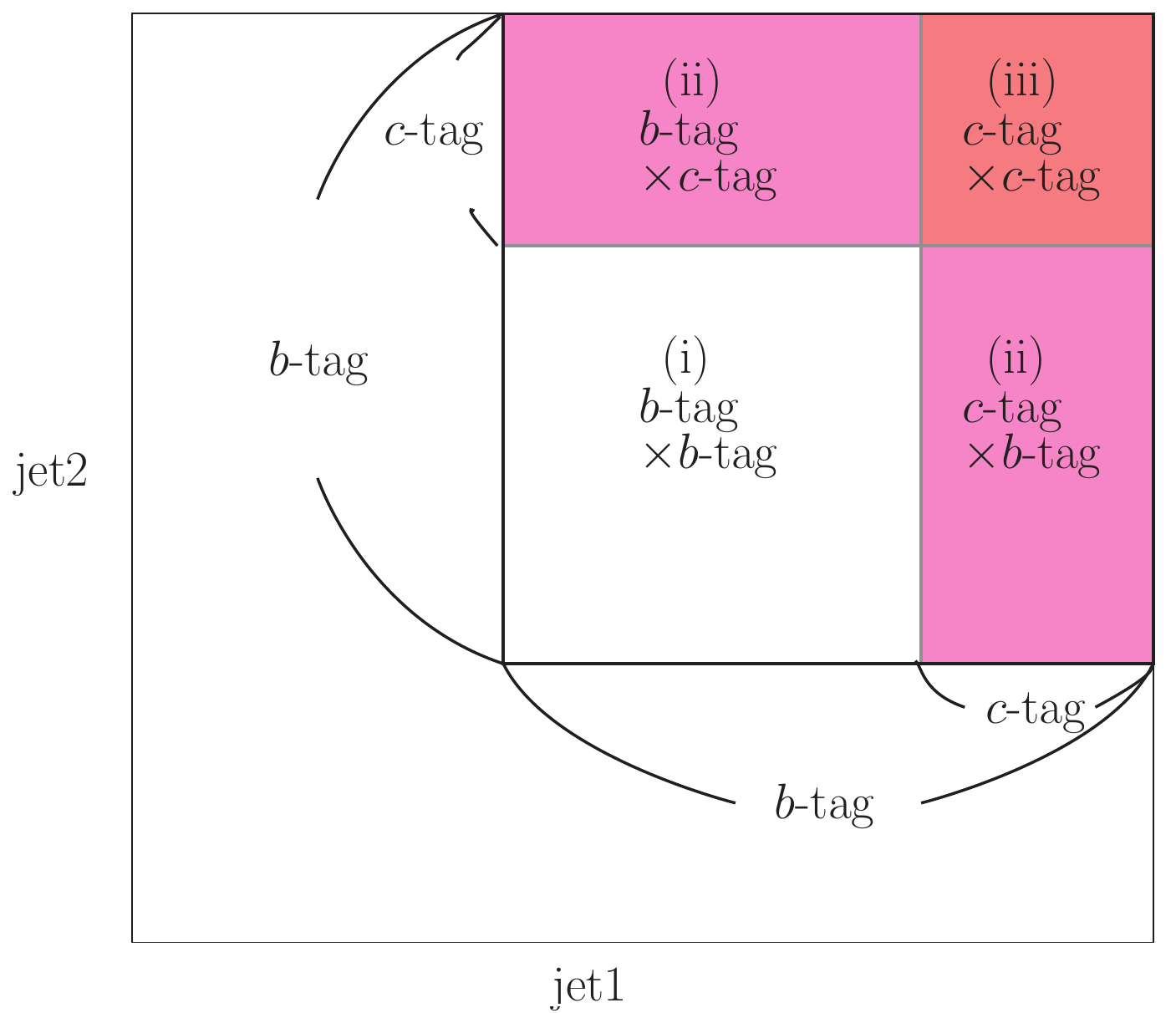}
\caption{Uncorrelated and correlated scenario in the left and right panel, respectively. 
$c$-tagged jets are a subset of $b$-tagged jets in the correlated scenario, 
while in the uncorrelated scenario $b$- and $c$-tagging cover different jets. 
\label{fig:scenarios}}
\end{figure}

$c$-tagging uses almost the same experimental information as $b$-tagging. 
As a result jets that are $c$-tagged may but also may not pass the $b$-tagging
criteria~\cite{ATL-PHYS-PUB-2015-001, Chatrchyan:2012jua}.
The actual experiments can employ $b$- and $c$-tagging simultaneously, but for 
our analysis it is not possible to fully take into account this correlation. 
Therefore, whenever possible, we study the following two extreme scenarios.
\begin{description}
\item[Uncorrelated scenario] 
$b$- and $c$-tagging are uncorrelated and possible to employ simultaneously. 
In this case, if a jet is $b$-tagged, the jet is never $c$-tagged, and vice versa, i.e.,
there is no overlap between $b$- and $c$-tagged jets. 
\item[Correlated scenario] $c$-tagging is fully correlated with $b$-tagging 
and is a tighter version of $b$-tagging. 
In this case, $c$-tagged jets are always also $b$-tagged, but the opposite 
is not necessary, i.e., $c$-tagged jets are a subset of $b$-tagged jets.  
\end{description}
The actual situation is expected to be something between the following two scenarios.
However, we will show in Section~\ref{LHC14inclusive} that 
the final results in the two scenarios are similar. We define three categories by combining taggers: 
(i)~two jets are $b$-tagged, 
(ii)~one is $b$-tagged and one is $c$-tagged, 
(iii)~two jets are $c$-tagged. 
To avoid double counting in categories (i) and (ii) of the correlated scenario, 
the $c$-tagged jets are removed from $b$-tagged jets. 
A schematic picture of the three categories for the scenarios is shown in 
Fig.~\ref{fig:scenarios}. 
In our analysis we use {\tt RunDec}~\cite{Chetyrkin:2000yt} to compute the quark 
masses at the Higgs mass using the inputs from PDG~\cite{PDG} finding
$m_s=53.2\,$MeV, $m_c =0.612\,$GeV and $m_b=2.78$\, GeV.

%%%%%%%%%%%%%%%%%%%%%%%%%%%%%%%%%%%%%%%%%%%%%%%%
\subsection{LHC 14\,TeV} \label{LHC14inclusive}
%%%%%%%%%%%%%%%%%%%%%%%%%%%%%%%%%%%%%%%%%%%%%%%%
For LHC run~II and HL-LHC, we base our study on the dedicated ATLAS analysis 
for the future measurement of $\mu_b$ based on the Higgs production associated with $W/Z$ bosons 
at LHC $14$\,TeV with $300\,\fb^{-1}$ and $3000\,\fb^{-1}$~\cite{ATL-PHYS-PUB-2014-011}.
The analysis requires two $b$-tagged jets, i.e., it employs a single tagger, which
is insufficient to disentangle~$\mu_b$ and~$\mu_c$.
To discuss the future sensitivities, we thus need to estimate the number of 
signal and background events in the three categories for the correlated and uncorrelated 
scenario once also $c$-tagging is employed.
We utilize the Monte-Carlo~(MC) studies presented in Figs.~3--6 of Ref.~\cite{ATL-PHYS-PUB-2014-011} in the following way. 
These figures provide the number of events in each bin for signal and for each background after 
applying all cuts and requiring two $b$-tagged jets. 
Let us consider a bin of signal or a specific background that 
originally has an $x$- and a $y$-jet, where  $x,y=b,c,l$ (real $b$-jet, $c$-jet, light-jet),
and the number provided is $N$.%
\footnote{For instance, $x=y=b$ for $h\to \bbbar$ 
signal and $t\bar{t}$ background; $x=y=c$ for $W\!+\!\ccbar$ background; $(x=b, y=l)$ for  
single top background.} 
We then obtain the number of events for categories (i)--(iii) in
uncorrelated and correlated scenarios as below.
\begin{align}
\intertext{\bf Uncorrelated scenario:}
	&N^{\rm (i)}   = N,
	&&N^{\rm (ii)}  =  \frac{\epsilon_{x}^{\text{($b$-tag)}} \epsilon_{y}^{\text{($c$-tag)}}+\epsilon_{x}^{\text{($c$-tag)}} \epsilon_{y}^{\text{($b$-tag)}}}{\epsilon_{x}^{\text{($b$-tag)}} \epsilon_{y}^{\text{($b$-tag)}}} N,
	&& N^{\rm (iii)} = \frac{\epsilon_{x}^{\text{($c$-tag)}} \epsilon_{y}^{\text{($c$-tag)}}}{\epsilon_{x}^{\text{($b$-tag)}} \epsilon_{y}^{\text{($b$-tag)}}}N\,,
	&\\
\intertext{\bf Correlated scenario:}
	&N^{\rm (i)}   = N - N^{\rm (ii)}- N^{\rm (iii)},
	&&N^{\rm (ii)}  
	= \frac{\epsilon_{x}^{\text{($b$-tag)}} \epsilon_{y}^{\text{($c$-tag)}}+\epsilon_{x}^{\text{($c$-tag)}} \epsilon_{y}^{\text{($b$-tag)}}}{\epsilon_{x}^{\text{($b$-tag)}} \epsilon_{y}^{\text{($b$-tag)}}} N - 2N^{\rm (iii)},
	&&N^{\rm (iii)} 
	=\frac{\epsilon_{x}^{\text{($c$-tag)}} \epsilon_{y}^{\text{($c$-tag)}}}{\epsilon_{x}^{\text{($b$-tag)}} \epsilon_{y}^{\text{($b$-tag)}}} N\,.
	&\label{eq:uncorrelated}
\end{align}
The rescaling is done on a bin-by-bin basis. 
After rescaling different background differently ($N\to B_{X}$ with $X$  denoting the type of background), 
we obtain the total background for each category, $B^{\rm (i, ii, iii)}$, by summing all backgrounds, 
$B^{\rm (i)}=\sum_{X}^{\text{all}} B_{X}^{\rm (i)}$ and analogously for~(ii) and~(iii). 
The expected signal for each category is straightforward, $N\to S$. 

%%%%%%%%%
\begin{figure}[t!]
\centering
\includegraphics[]{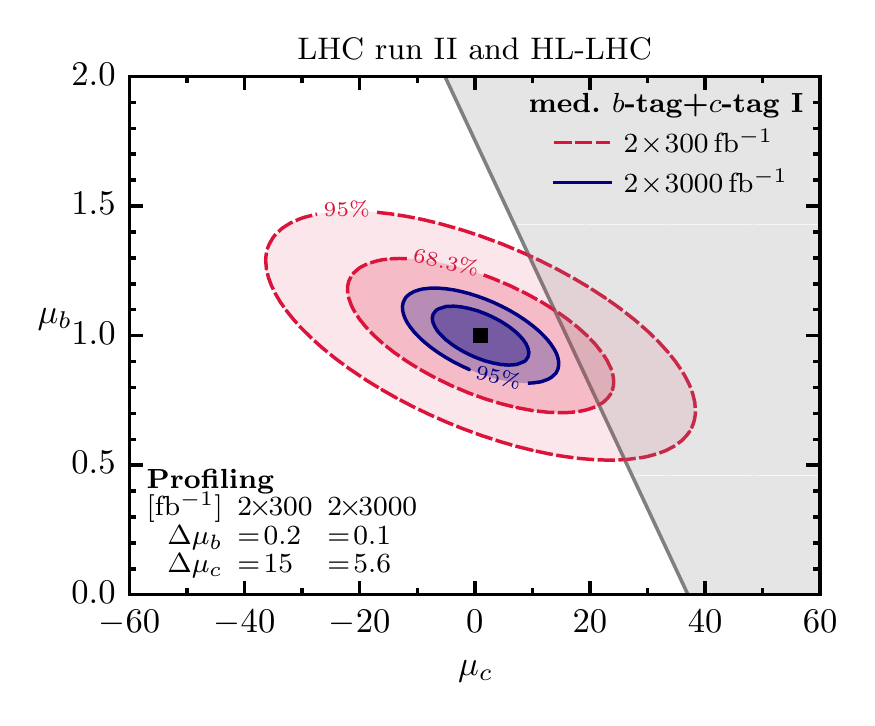}\hspace*{-1em}
\includegraphics[]{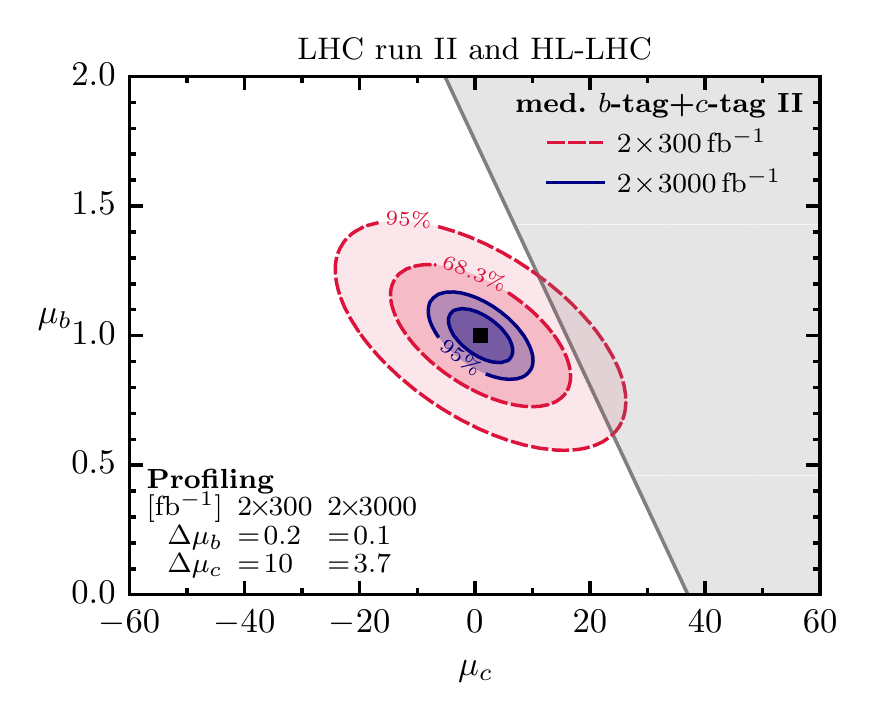}
\caption{$300$\,fb$^{-1}$ and $3000$\,fb$^{-1}$ prospects for the signal strengths at the LHC, for 
$h\to b\bar b$ and $h\to c\bar c$ based on $b$- and $c$-tagging for the uncorrelated scenario 
employing $c$-tagging I (left panel) and $c$-tagging II (right panel).
The grey shaded region is unphysical unless Higgs production is modified with respect to
the SM case. The profiled likelihood ratio \cite{Cowan:2010js} is used for the respective 1-$\sigma$ uncertainty of $\Delta\mu_b$ and $\Delta\mu_c$. 
\label{fig:mub-muc}}
\end{figure}
%%%%%%%%%

To obtain the future sensitivity we then follow the statistical procedure described in 
Ref.~\cite{Perez:2015aoa}. 
Given the expectation of signal and background, we construct a likelihood function of 
$\mu_c$ and $\mu_b$ based on the Poisson probability-distribution function, and use  the likelihood ratio for parameter estimates. 
In Fig.~\ref{fig:mub-muc} we present the future reach  for the signal strengths 
of $h\to b\bar b$ and $h\to c\bar c$ in the uncorrelated scenario by combining 
$b$-tagging with $c$-tagging~I~(left panel) and~II~(right panel).
Note that the correlated scenario cannot be defined in the case of $c$-tagging~II and~III, 
see Appendix~\ref{app:inclusive} for details and Fig.~\ref{fig:mub-muc-corr} therein for the 
$c$-tagging~I result of the correlated scenario. 
We obtain the expected uncertainty on $\mu_c$ ($\mu_b$) by profiling $\mu_b$ ($\mu_c$).
We list the $1$-$\sigma$ ranges for $\mu_c$ and $\mu_b$ for different scenarios 
and employed $c$-tagging assuming the total luminosity of $2\times300$\,fb$^{-1}$ and 
$2\times3000$\,fb$^{-1}$ expected at LHC run~II and HL-LHC
in Tab.~\ref{tab:profiles14TeV}.
The sensitivities for $c$-tagging I in the correlated and uncorrelated scenario are similar, so
we conclude that these results represent well the actual future reach.

\begin{table}[t]
\begin{tabular}[t]{c@{\hskip 1.5em}l@{\hskip 1.5em}c@{\hskip 1.5em}c@{\hskip 1.5em}c@{\hskip 1.5em}c@{\hskip 1.5em}c@{\hskip 1.5em}c}
					$\cal L$	
					& \multicolumn{1}{c}{$\sqrt{s}=14$\,TeV}
					& $\Delta\mu_b$	
					& $\Delta\mu_c$	
					& $\Delta\kappa_b$	
					& $\Delta\kappa_c$	
					& $\kappa_b$ @ $95\%$\,CL	
					& $|\kappa_c|$ @ $95\%$\,CL	
					\\\hline\hline
$2\times300$\,fb$^{-1}$  
                  &{ ~~\;\,correlated $c$-tagging I}	& $\pm0.22$ & $\pm15$  & $+1.3$  & $+8.6$ & $[0.67,7.07]$ & $<37$\\
                  &{ uncorrelated $c$-tagging I}	& $\pm0.20$ & $\pm15$  & $+1.5$  & $+9.4$ & $[0.69,7.16]$ & $<38$\\
                  &{ uncorrelated $c$-tagging II}	& $\pm0.18$ & $\pm10$  & $+0.5$  & $+4.1$ & $[0.70,4.70]$ & $<21$\\
                  &{ uncorrelated $c$-tagging III}	& $\pm0.17$ & $\pm5.8$ & $+0.29$ & $+2.2$ & $[0.70,1.90]$ & $<6.0$\\
		  \hline
$2\times3000$\,fb$^{-1}$ 
                  &{ ~~\;\,correlated $c$-tagging I}	& $\pm0.084$ & $\pm5.6$ & $+0.20$ & $+2.1$ & $[0.84,1.57]$ & $<5.5$\\
                  &{ uncorrelated $c$-tagging I}	& $\pm0.075$ & $\pm5.6$ & $+0.20$ & $+2.1$ & $[0.85,1.60]$ & $<5.6$\\
                  &{ uncorrelated $c$-tagging II}	& $\pm0.069$ & $\pm3.7$ & $+0.12$ & $+1.4$ & $[0.86,1.30]$ & $<3.7$\\
                  &{ uncorrelated $c$-tagging III}	& $\pm0.065$ & $\pm2.0$ & $+0.087$ & $+0.82$ & $[0.87,1.18]$ & $<2.5$\\
		  \hline
\end{tabular}
\caption{$1$-$\sigma$ uncertainties after profiling \cite{Cowan:2010js} for the signal strengths, $\mu_b$ and $\mu_c$, and the
Yukawa-coupling modifications, $\kappa_b$ and $\kappa_c$ at the LHC with  $\sqrt{s}=14$\,TeV. 
The results for different scenarios (uncorrelated and correlated) and different $c$-tagging are
shown for a total luminosity of  $2\times300$\,fb$^{-1}$ and $2\times3000$\,fb$^{-1}$.
The error in $\kappa_b$ and $\kappa_c$ is asymmetric; we only show the upper
bound. The $95\%$\,CL regions for $\kappa_b$ and $\kappa_c$ are given in the
last two columns.
\label{tab:profiles14TeV}}
\end{table}

The translation of the constraints of the charm and bottom signal strengths 
to the Yukawa couplings themselves requires some caution. 
If we assume that Higgs production is not modified with respect to the SM,
the signal strengths are given by $\mu_c = \BR_\ccbar/\BR^{\rm SM}_\ccbar$ and 
$\mu_b = \BR_\bbbar/\BR^{\rm SM}_\bbbar$. 
In the extreme case in which the Higgs decays solely to charms and bottoms, $\BR_\ccbar+\BR_\bbbar=1\,$ 
holds and the two rates are linearly dependent. 
As long as the measured values of $\mu_{c}$ and $\mu_b$ are consistent with this hypothesis,
an arbitrary large value of $\kappa_c\equiv y_c/y_c^{\rm SM}$ is allowed with some 
$\kappa_b\equiv y_b/y_b^{\rm SM}$. This corresponds to a flat direction in the 
$\kappa_c$--$\kappa_b$ plane. 
In other words, as long as the experimental result is consistent with
\begin{align}
	\mu_c \BR^{\rm SM}_\ccbar + \mu_b \BR^{\rm SM}_\bbbar > 1 \, ,
\end{align}
one cannot constrain $\kappa_c$ and $\kappa_b$ assuming only SM Higgs production. 
We illustrate this case with the grey shaded region in Fig.~\ref{fig:mub-muc}. 
If this region overlaps with the allowed regions of $\mu_c$--$\mu_b$ (coloured ellipses), 
it means that there is a flat direction in the $\kappa_c$--$\kappa_b$ if SM Higgs production
is assumed.

A charm Yukawa much larger than in the SM enhances the Higgs production in the $Vh$ 
production channel; for $\kappa_c\!\sim\!\cO(100)$ it is twice as large as the SM expectation~\cite{Perez:2015aoa}. 
This mechanism enabled us to obtain a direct constraint on $\kappa_c$ already with the available 
$8$\,TeV dataset~\cite{Perez:2015aoa}.
For the $14$\,TeV projection, as seen in Fig.~\ref{fig:mub-muc}, considering non-SM production 
is essential to constrain $\kappa_c$ with $300\,\fb^{-1}$, while for the high-luminosity 
stage its effect is minor. Details of non-SM $Vh$ production at 14\, TeV are discussed in Appendix~\ref{app:inclusive}.

\begin{figure}[t!]
\centering
\includegraphics[]{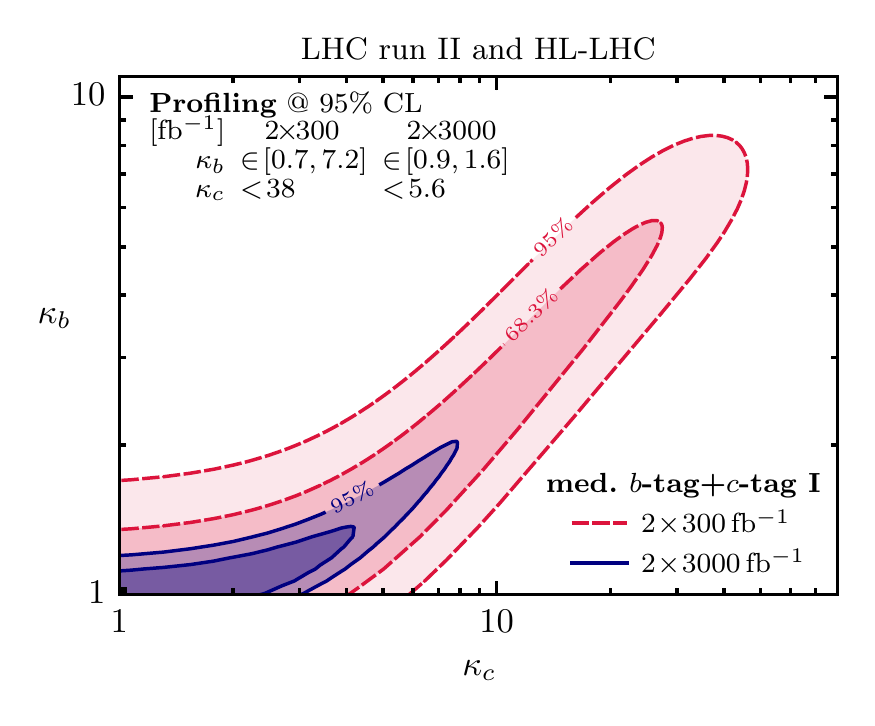}\hspace*{-1em}
\includegraphics[]{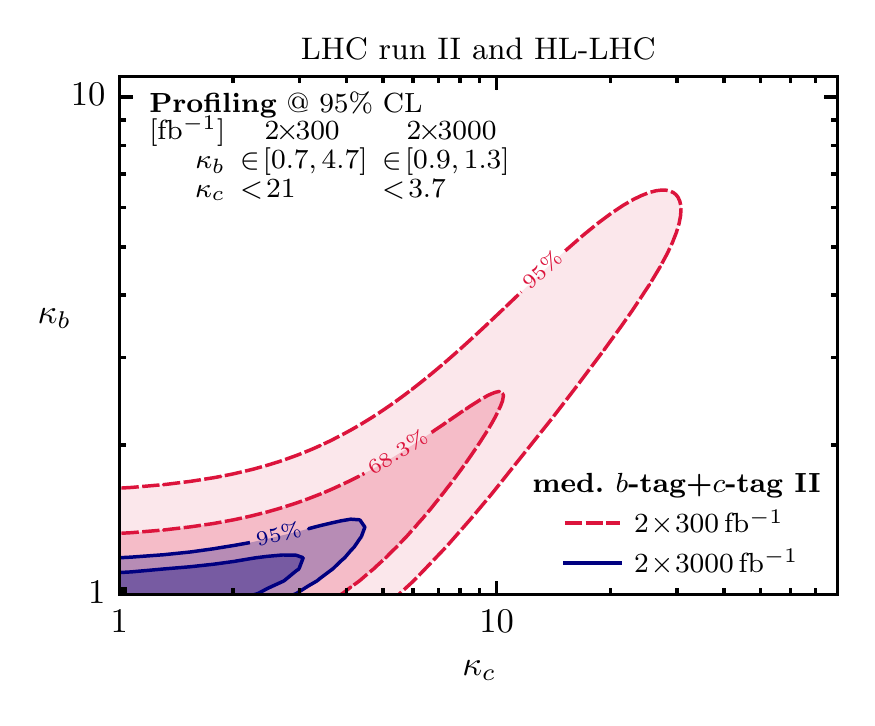}
\caption{$300$\,fb$^{-1}$ and $3000$\,fb$^{-1}$ prospects for probing 
$\kappa_b$ and $\kappa_c$ at the LHC, with $h\to b\bar b$ and $h\to c\bar c$ based 
on $b$- and $c$-tagging for the uncorrelated scenario employing $c$-tagging I (left panel) and $c$-tagging II
(right panel).
All other Higgs couplings are assumed to be like in the SM. 
The profiled likelihood ratio \cite{Cowan:2010js} is used for the respective reach on $\kappa_b$ and $\kappa_c$.
\label{fig:kab-kac}}
\end{figure}

In the analysis for the prospects for couplings, we float only $\kappa_b$ and $\kappa_c$ freely and assume the other couplings stay as in the SM, in particular, $\kappa_V=1$. 
In Fig.~\ref{fig:kab-kac} we show the expected future reach in the $\kappa_c$--$\kappa_b$ plane 
taking into account non-SM $Vh$ production for the uncorrelated scenario employing $c$-tagging I and
II. 
We obtain the expected upper bound on $\kappa_c$ ($\kappa_b$) by profiling over $\kappa_b$ ($\kappa_c)$ \cite{Cowan:2010js}. 
The $1$-$\sigma$ uncertainties, as well as the $95\%$\,CL ranges for $\kappa_b$
and $\kappa_c$ for the different cases are listed in Tab.~\ref{tab:profiles14TeV}.

Finally, we compare the projected reach of the direct charm-Yukawa measurement
to the indirect bound from the global analysis of the Higgs couplings. 
For $\kappa_c<5$ the effects of non-SM Higgs production due to large charm Yukawa are 
negligible and the constraint on $\kappa_c$ can be deduced from the bound on the untagged 
Higgs decays. 
The ATLAS projection for the HL-LHC is $\BR_{\rm untagged}<10\,\%$ at $95\%$\,CL 
(without theoretical uncertainties)~\cite{ATL-PHYS-PUB-2014-016}. 
It can be interpreted as $\kappa_c\lesssim2\,$. 
This upper bound is comparable to our projection with $c$-tagging~III, 
see Tab.~\ref{tab:profiles100TeVboosted}.

%%%%%%%%%
%%%%%%%%%

%%%%%%%%%%%%%%%%%%%%%%%%%%%%%%%%%%%%%%%%%%%%%%%%
\subsection{$pp$ collider with $\sqrt{s}=100$\,TeV}
In this section we perform a first study of the sensitivity reach of a $100$\,TeV $pp$ collider
in measuring the charm-quark Yukawa via the inclusive rate.
As a byproduct we obtain also the sensitivity of a bottom-Yukawa measurement
at $100$\,TeV. 
For such a machine, there exists no detailed, fully realistic study for the $h\to b \bar b$ 
prospects like the one of ATLAS for $14$\,TeV \cite{ATL-PHYS-PUB-2014-011}, which
we employed for our $14$\,TeV study.
For this reason, we investigate the $100$\,TeV reach by simulating the signal and
main backgrounds at the leading-order (LO) parton level using {\tt MadGraph} 5.2~\cite{Alwall:2011uj} and multiplying with the
inclusive $k$-factors. 

The main difficulty remains to find a way to reduce background, while keeping as many
signal $h\to c\bar c$ events as possible.
To this end, we follow two orthogonal directions. 
Firstly, we look into the boosted-Higgs regime in which the Higgs has $p_T(h)>350$\,GeV. 
In this case we rely on available jet-substructure techniques to extract the signal
and reduce the $t\bar t$ background that dominates in this kinematic
configuration.
Secondly, we ``unboost'' the Higgs by binning in $H_T$. 
This way the $S/B$ ratio for $h\to c\bar c$ is large in lower $H_T$ bins as the main background, $t\bar t$,
typically has higher $H_T$ than the signal.

We shall find that the sensitivity reach for the bottom Yukawa is not
significantly different in the two cases, as in both there are enough
$h\to b\bar b$ events.
For the charm Yukawa, however, the ``unboosted'' analysis appears more 
promising, due to the fact that it accepts a larger fraction of the rather rare signal events.
Given that the capabilities of a future $100$\,TeV collider and the advancements 
with respect to current experiments are currently not well known, the fact that our 
projections will be based on LO simulations suffices.
However, it is important to note that we expect significantly better results from 
realistic studies that employ multiple bins with increased $S/B$ ratio.
For instance, the projected uncertainty on $\Delta \mu_b$ from Ref.~\cite{ATL-PHYS-PUB-2014-011} would
be approximately a factor of $2$ larger without binning.
Furthermore, in Ref.~\cite{ATL-PHYS-PUB-2014-011} the sensitivity of a purely cut-based
analysis was compared to the one obtained employing multivariate techniques.
In the latter the uncertainty is decreased by roughly $25\%$.
This gives us confidence that the results presented here are conservative and
there is room for improvements in the future.

\subsubsection*{Boosted-Higgs analysis}
The field of searching for boosted massive particles and jet-substructure is very rich 
and we shall not attempt to describe it here in any detail 
(see {\it e.g.}~\cite{Altheimer:2013yza} for a recent review).
Instead we focus on one specific method to study the sensitivity to the  
Higgs couplings to bottom and charm quarks at a $100$\,TeV collider.
The sensitivity to the $h\to b\bar b$ decay mode with the Higgs being boosted and
produced in association with a leptonically decaying $W$ at the LHC with $\sqrt{s}=8$ and $13$\,TeV 
has been analysed in Ref.~\cite{Backovic:2012jj}. The study adopted the Template Overlap 
Method~\cite{Almeida:2010pa,Almeida:2011aa} (see also~\cite{Aad:2012raa} for the ATLAS implementation 
of the method).
For our study we will use the signal efficiency and background-rejection rates of 
the ``Cuts 5'' scenario in Ref.~\cite{Backovic:2012jj} and a cut on the fat jet containing the Higgs (or its $b\bar b$ daughter products), $p_T(h)>350$\,GeV. 
Given the above requirements, the $Wh$ signal has an efficiency of $22\%$ 
while the 
$t\bar t$ and $Wb\bar b$ backgrounds have a fake rate of only
$1.3\%$ and $5.1\%$, respectively (see Tab.~III in Ref.~\cite{Backovic:2012jj}).
We will assume that these jet-substructure efficiencies do not change from $13$ to $100$\,TeV.

%%%%%%%%%
\begin{figure}[t!]
\centering
\includegraphics[]{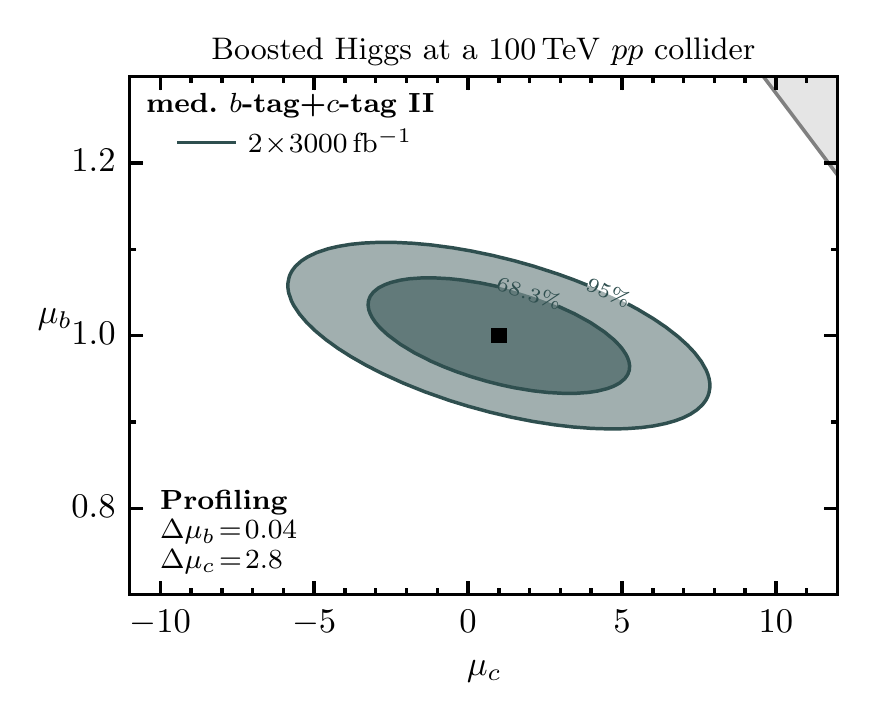}\hspace*{-1em}
\includegraphics[]{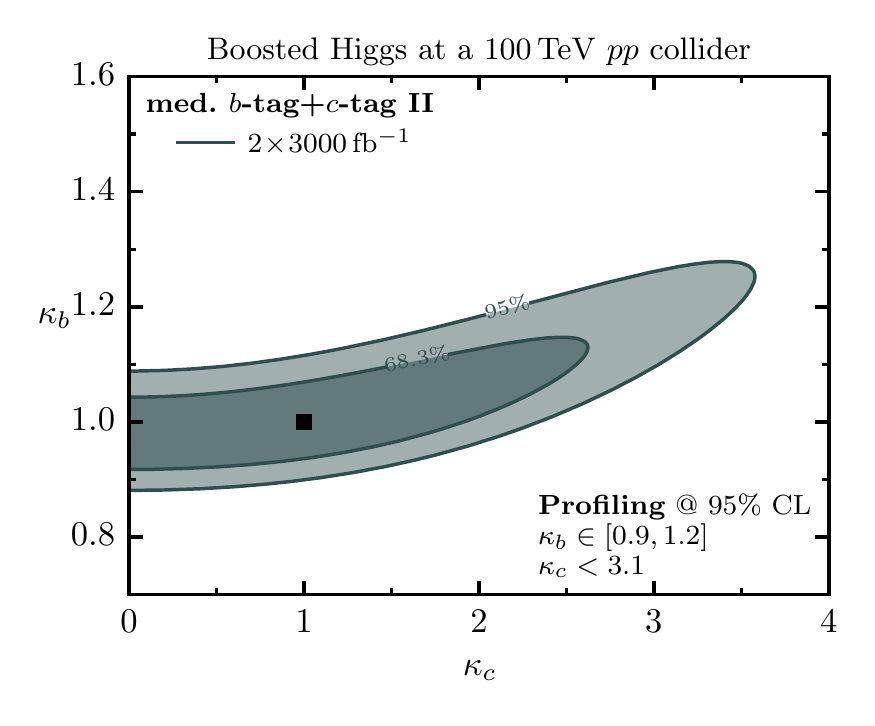}
\caption{Prospects for probing $\mu_b$, $\mu_c$ (left panel), 
and $\kappa_b$, $\kappa_c$ (right panel) at a $100$\,TeV collider, with $h\to b\bar b$ and $h\to c\bar c$ based on 
$b$- and $c$-tagging II for the uncorrelated scenario using boosted Higgses
and $2\times3000$\,fb$^{-1}$ of data.
The profiled likelihood ratio \cite{Cowan:2010js} is used for the respective reach of $\Delta\mu_b$, $\Delta\mu_c$, $\kappa_b$, and $\kappa_c$.
\label{fig:100TeVboosted}}
\end{figure}
%%%%%%%%%

To make use of these jet-substructure results for our $100$\,TeV study
we follow their analysis and simulate signal and background for both $13$ and 
$100$\,TeV applying the same basic cuts. 
The main requirement is the presence of two $b$-tagged jets inside the fat jet
and a few basic cuts the most relevant of which are $p_T(W), p_T({\text{fat jet}}) > 350$\,GeV 
and $0.4<\Delta R_{bb}<0.8$ (see Eq.~(12) and (13) in Ref.~\cite{Backovic:2012jj}). 
Their simulation of the signal $Wh$, and the backgrounds $Wb\bar b$, $t\bar t$ 
includes matching to parton shower and next-to-leading-order (NLO) $k$-factors from 
{\tt MCFM~6.3}~\cite{Campbell:2010ff}. 
We include these NLO effects by rescaling our LO parton-level simulation at $13$\,TeV
to their results and applying the same rescaling factors to the $100$\,TeV results.
In a similar way, we also include in our study the two-lepton sample, namely 
$Zh$ production with leptonically decaying $Z$ and the dominant corresponding backgrounds 
$Zb\bar b$ and leptonic $t\bar t$. 
We use the same rescaling factors as for the $Wh$ sample.
For a charm-Yukawa measurement it is necessary to include the $Wc\bar c$ and $Zc\bar c$
backgrounds, because they can be relevant when we employ $c$-tagging with a
large tagging efficiency for charm quarks. 
The rescaling factor from $Wb\bar b$ is used for both of them to rescale their $100$\,TeV 
cross sections.
Finally, we note that the main $t\bar t$ background in the one-lepton analysis
originates from a fat jet consisting of a bottom quark and a mistagged charm quark from the 
associated hadronically decaying $W$ \cite{Backovic:2012jj}.
Such a configuration is absent in the two-lepton, $Zh$, sample, which has thus 
a reduced $t\bar t$ background.

Having simulated the signal and the dominant backgrounds at $100$\,TeV and
rescaled with the $13$\,TeV $k$-factors we find the expected signal and background
events and multiply with the corresponding efficiencies depending on whether $b$-tagging 
or $c$-tagging I, II, III is applied.
The rest of the analysis is analogous to the $14$\,TeV study. 
Also here we combine $b$-tagging with one $c$-tagging scenario and profile the resulting distributions to obtain the 
future sensitivity in signal strengths and Yukawa-coupling modifications.
Here, we present results assuming a total luminosity of $3000$\,fb$^{-1}$ and  $2\times3000$\,fb$^{-1}$.
In Fig.~\ref{fig:100TeVboosted} we show the result of combining $b$- and $c$-tagging II
for the uncorrelated scenario using $2\times3000$\,fb$^{-1}$ of data. The
expected signal-strength (Yukawa-coupling) regions of sensitivity are plotted in 
the left (right) panel.
The $95\%$\,CL region has no overlap with the shaded grey region that is unphysical
if SM-Higgs production is assumed.
Therefore, modifications in the Higgs production are small and we can safely
neglect them.
In Tab.~\ref{tab:profiles100TeVboosted} we list the projected $1$-$\sigma$
uncertainties for $\mu_b$, $\mu_c$ and $\kappa_b$, $\kappa_c$ for various
scenarios.
For the Yukawa coupling modification we also present the $95\%$\,CL region after profiling.

Regarding $\mu_b$, we find that the expected precision is better in the correlated
scenario than in the uncorrelated one, see Tab.~\ref{tab:profiles100TeVboosted}.
The reason is that, as discussed, the main background in this boosted regime is
$t\bar t$ with a mistagged $c$ quark.
The separation of these background events with the correlated prescription of
Eq.~\eqref{eq:uncorrelated} assigns most of them to category (ii).
This results in an increased $S/B$ ratio in category (i) and leads to a better
expected precision in $\mu_b$ than in the corresponding uncorrelated case.
As far as $\mu_c$ is concerned, we find moderate improvements in sensitivity with respect to HL-LHC, 
compare the results in Tab.~\ref{tab:profiles14TeV} with those in
Tab.~\ref{tab:profiles100TeVboosted}. For instance for $2\times 3000$\,fb$^{-1}$
using $c$-tagging~II the improvement in the sensitivity for $\mu_c$ is approximately
$24\%$.
The reason for this is that, even though the jet-substructure cuts remove a
lot of background, a lot of $h\to c\bar c$ signal is also lost in this boosted regime.
We, therefore, look into the orthogonal direction of ``unboosting'' the Higgs.

\begin{table}[t]
\begin{tabular}[t]{c@{\hskip 1.5em}l@{\hskip 1.5em}c@{\hskip 1.5em}c@{\hskip 1.5em}c@{\hskip 1.5em}c@{\hskip 1.5em}c@{\hskip 1.5em}c}
					$\cal L$	
					& \multicolumn{1}{c}{boosted $\sqrt{s}=100$\,TeV}
					& $\Delta\mu_b$	
					& $\Delta\mu_c$	
					& $\Delta\kappa_b$	
					& $\Delta\kappa_c$	
					& $\kappa_b$ @ $95\%$\,CL	
					& $|\kappa_c|$ @ $95\%$\,CL	
					\\\hline\hline
$3000$\,fb$^{-1}$  
                  &{ ~~\;\,correlated $c$-tagging I}	& $\pm0.048$ & $\pm5.7$ & $+0.22$ & $+2.2$ & $[0.89,1.72]$   & $<6.1$\\
                  &{ uncorrelated $c$-tagging I}	& $\pm0.067$ & $\pm5.8$ & $+0.22$ & $+2.2$ & $[0.86,1.71]$   & $<6.1$\\
                  &{ uncorrelated $c$-tagging II}	& $\pm0.063$ & $\pm4.0$ & $+0.14$ & $+1.5$ & $[0.87,1.34]$   & $<4.0$\\
                  &{ uncorrelated $c$-tagging III}	& $\pm0.059$ & $\pm2.2$ & $+0.089$ & $+0.90$ & $[0.87,1.19]$ & $<2.6$\\
		  \hline
$2\times3000$\,fb$^{-1}$ 
                  &{ ~~\;\,correlated $c$-tagging I}	&$\pm0.034$ & $\pm4.0$ & $+0.14$ & $+1.6$ & $[0.92,1.36]$   & $<4.1$\\
                  &{ uncorrelated $c$-tagging I}	&$\pm0.048$ & $\pm4.1$ & $+0.14$ & $+1.6$ & $[0.89,1.36]$   & $<4.1$\\
                  &{ uncorrelated $c$-tagging II}	&$\pm0.044$ & $\pm2.8$ & $+0.090$ & $+1.1$ & $[0.90,1.20]$  & $<3.1$\\
                  &{ uncorrelated $c$-tagging III}	&$\pm0.042$ & $\pm1.6$ & $+0.061$ & $+0.67$ & $[0.90,1.13]$ & $<2.2$\\
		  \hline
\end{tabular}
\caption{$1$-$\sigma$ uncertainties after profiling \cite{Cowan:2010js} for the signal strengths, $\mu_b$ and $\mu_c$, and the
Yukawa-coupling modifications, $\kappa_b$ and $\kappa_c$, for a boosted Higgs using jet substructure 
at a $pp$ collider with  $\sqrt{s}=100$\,TeV. 
The results for different scenarios (uncorrelated and correlated) and different $c$-tagging are
shown for a total luminosity of  $3000$\,fb$^{-1}$ and $2\times3000$\,fb$^{-1}$.
The $95\%$\,CL regions for $\kappa_b$ and $\kappa_c$ are given in the
last two columns.
\label{tab:profiles100TeVboosted}}
\end{table}

\subsubsection*{``Unboosted''-Higgs analysis}
Our ``unboosted'' $100$\,TeV analysis is conceptually not much different than the
$14$\,TeV projection analysis of ATLAS~\cite{ATL-PHYS-PUB-2014-011}. 
Unlike the previous analysis, here we also include less energetic events in which the Higgs is not necessarily boosted. 
Relaxing the requirement for a large boost increases the otherwise statistically challenged 
$h\to c\bar{c}$ signal. We therefore expect an improved sensitivity for  
$\mu_c$ without affecting much the $\mu_b$ sensitivity.   

%%%%%%%%%
\begin{figure}[t!]
\centering
\includegraphics[]{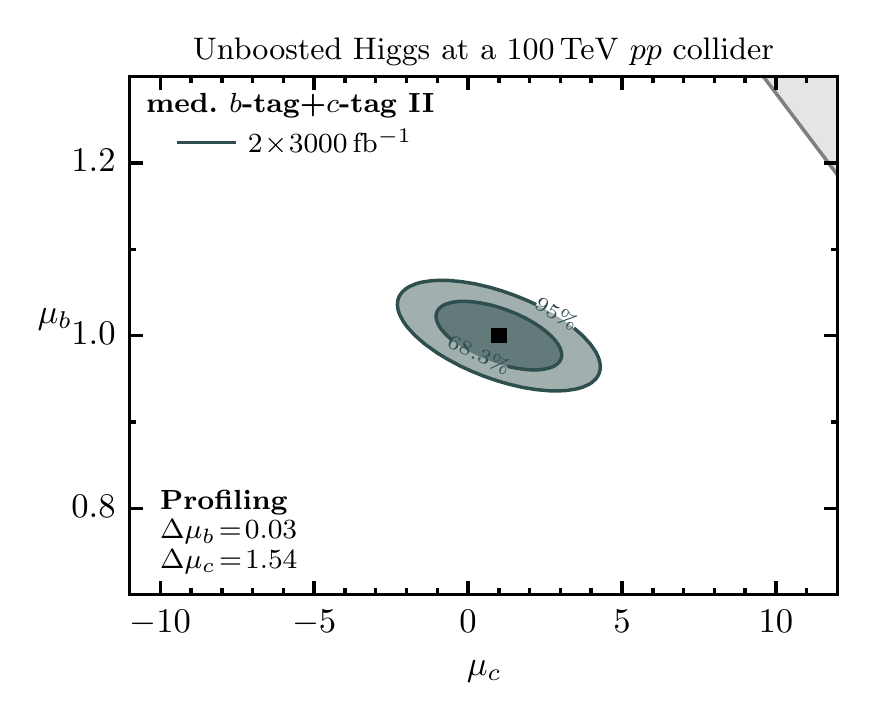}\hspace*{-1em}
\includegraphics[]{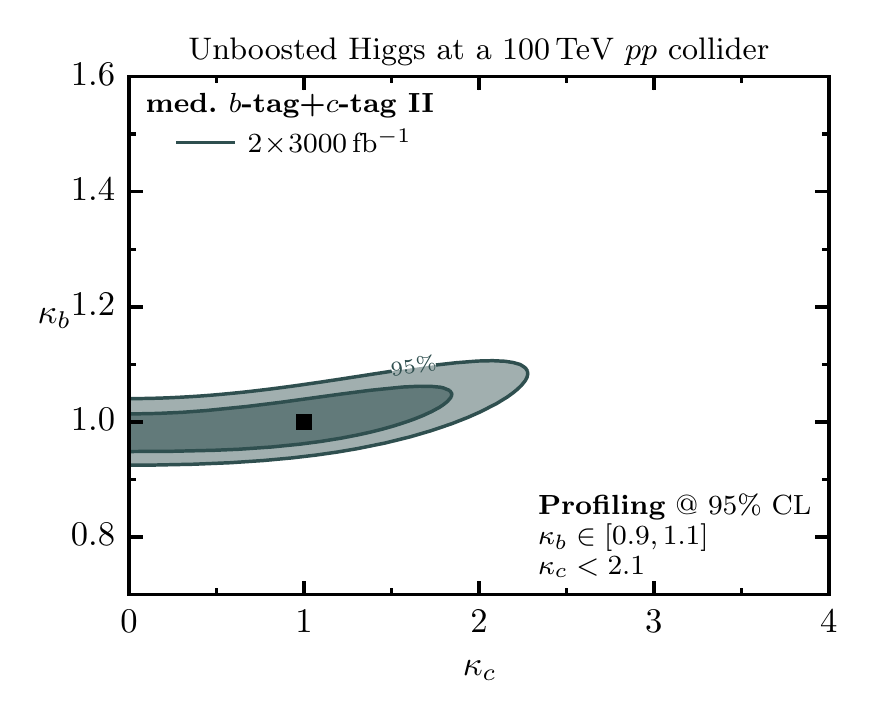}
\caption{Prospects for probing $\mu_b$, $\mu_c$ (left panel), 
and $\kappa_b$, $\kappa_c$ (right panel) at a $100$\,TeV collider, with $h\to b\bar b$ and $h\to c\bar c$ based on 
$b$- and $c$-tagging II for the uncorrelated scenario using non-boosted Higgses
and $2\times3000$\,fb$^{-1}$ of data.
The profiled likelihood ratio \cite{Cowan:2010js} is used for the respective reach of $\Delta\mu_b$, $\Delta\mu_c$, $\kappa_b$ and $\kappa_c$.
\label{fig:100TeVunboosted}}
\end{figure}
%%%%%%%%%

We will use three bins of inclusive $H_T$, 
\begin{align}
	H_T<340\, {\rm GeV},  \quad	340\, {\rm GeV}<H_T<500\, {\rm GeV}, \quad 500\, {\rm GeV}<H_T\ .
\end{align}
In the lower $H_T$ bins the $t\bar t$ background is reduced.
The main basic cuts that we apply are: $p_T(W/Z) > 100$\,GeV,
$p_T(j_1) > 60$\,GeV, $p_T(j_2) > 40$\,GeV, $\Delta R(j_1,j_2)>0.4$ and 
$100$\,GeV$<m_{j_1j_2}<140$\,GeV, where $j_1$ and $j_2$ is the leading 
and next-to-leading in $p_T$ jet, respectively.
Following Ref.~\cite{ATL-PHYS-PUB-2014-011} we demand for the one-lepton sample 
$E_T^{\rm miss}>40$\,GeV and for the two-lepton channel $E_T^{\rm miss}<60$\,GeV.
We simulate the same background processes as for the boosted $100$\,TeV analysis
and rely on LO parton-level simulation supplemented with the same inclusive
$k$-factors.
A difference with respect to the boosted analysis is that the $t\bar t$ background
in this case is not completely dominated by the mistagged $c$-quark, i.e.\ the
background from two $b$-quarks is at least as equally important. 
The ratio of these two backgrounds depends on the jet-tagging employed, so 
in our study, we keep them as separate backgrounds that depend differently on
$\epsilon_b$ and $\epsilon_c$.

\begin{table}[t]
\begin{tabular}[t]{c@{\hskip 1.5em}l@{\hskip 1.5em}c@{\hskip 1.5em}c@{\hskip 1.5em}c@{\hskip 1.5em}c@{\hskip 1.5em}c@{\hskip 1.5em}c}
					$\cal L$	
					& \multicolumn{1}{c}{unboosted $\sqrt{s}=100$\,TeV}
					& $\Delta\mu_b$	
					& $\Delta\mu_c$	
					& $\Delta\kappa_b$	
					& $\Delta\kappa_c$	
					& $\kappa_b$ @ $95\%$\,CL	
					& $|\kappa_c|$ @ $95\%$\,CL	
					\\\hline\hline
$3000$\,fb$^{-1}$  
                  &{ ~~\;\,correlated $c$-tagging I}	&$\pm0.047$ & $\pm3.2$ & $+0.086$ & $+1.1$  & $[0.91,1.20]$ & $<3.0$\\
                  &{ uncorrelated $c$-tagging I}	&$\pm0.047$ & $\pm3.2$ & $+0.086$ & $+1.1$  & $[0.91,1.20]$ & $<3.0$\\
                  &{ uncorrelated $c$-tagging II}	&$\pm0.044$ & $\pm2.2$ & $+0.057$ & $+0.80$ & $[0.92,1.12]$ & $<2.4$\\
                  &{ uncorrelated $c$-tagging III}	&$\pm0.041$ & $\pm1.1$ & $+0.043$ & $+0.46$ & $[0.92,1.09]$ & $<1.8$\\
		  \hline
$2\times3000$\,fb$^{-1}$ 
                  &{ ~~\;\,correlated $c$-tagging I}	&$\pm0.034$ & $\pm2.3$ & $+0.058$ & $+0.83$ & $[0.93,1.13]$ & $<2.5$\\
                  &{ uncorrelated $c$-tagging I}	&$\pm0.034$ & $\pm2.3$ & $+0.058$ & $+0.83$ & $[0.93,1.13]$ & $<2.5$\\
                  &{ uncorrelated $c$-tagging II}	&$\pm0.031$ & $\pm1.5$ & $+0.039$ & $+0.59$ & $[0.94,1.08]$ & $<2.1$\\
                  &{ uncorrelated $c$-tagging III}	&$\pm0.029$ & $\pm0.8$ & $+0.030$ & $+0.34$ & $[0.95,1.06]$ & $<1.6$\\
		  \hline
\end{tabular}
\caption{$1$-$\sigma$ uncertainties after profiling \cite{Cowan:2010js} for the signal strengths, $\mu_b$ and $\mu_c$, and the
Yukawa-coupling modifications, $\kappa_b$ and $\kappa_c$, for non-boosted Higgses 
at a $pp$ collider with  $\sqrt{s}=100$\,TeV. 
The results for
different scenarios (uncorrelated and correlated) and different $c$-tagging are
shown for a total luminosity of  $3000$\,fb$^{-1}$ and $2\times3000$\,fb$^{-1}$.
The $95\%$\,CL regions for $\kappa_b$ and $\kappa_c$ are given in the
last two columns.
\label{tab:profiles100TeVunboosted}}
\end{table}

Having simulated the $Wh$ and $Zh$ signal processes and the corresponding
background processes for the three $H_T$ bins we find the projected sensitivities
by combining all bins using $b$-tagging together with one $c$-tagging scenario.
The details are analogous to the boosted analysis.
In Tab.~\ref{tab:profiles100TeVunboosted} we list the projected $1$-$\sigma$ uncertainties 
for signal strengths and Yukawa-coupling modifications as well as the $95\%$\,CL region
for the latter assuming $3000$\,fb$^{-1}$ and $2\times3000$\,fb$^{-1}$.
In Fig.~\ref{fig:100TeVunboosted} we present the expected signal-strength and 
Yukawa-coupling sensitivity regions when $c$-tagging~II is employed and 
$2\times3000$\,fb$^{-1}$ of data are assumed. 
In this case, we find for the signal strength $\mu_c$ a significant improvement
of approximately $60\%$ in the projected uncertainty with respect to the analogous 
$14$\,TeV expectation.
We also find the promising result that, if $c$-tagging~III is possible, a factor of 
$\sim2$ modification in the charm-quark Yukawa will be probed by a $100$\,TeV collider with $95\%$\,CL.

%%%%%%%%%%%%%%%%%%%%%%%%%%%%%%%%%%%%%%%%%%%%%%%%
\subsection{$e^+e^-$ colliders} \label{sec:ee}
%%%%%%%%%%%%%%%%%%%%%%%%%%%%%%%%%%%%%%%%%%%%%%%%
Since the future $e^+e^-$ colliders will be experiments aiming at Higgs precision, 
we expect significant improvement in measuring Higgs couplings. 
Here, we show the summary of their prospects. There are two types of  $e^+e^-$ 
colliders proposed, linear type, such as ILC, and circular type, such as TLEP. 
The advantage of the ILC is that it relies on established technology and its collision 
energy can be potentially increased up to $1$\,TeV, while in the TLEP one expects 
much larger luminosity by an order of magnitude and so many Higgs events will be collected.   

The technical design report of ILC~\cite{Baer:2013cma} presents the expected precision 
in different channels based on dedicated analyses, 
\begin{align}
	\Delta \mu_{b}= 1.1\%\ (0.47\%) \, ,  \quad
	\Delta \mu_{c}=7.4\%\ (6.7\%),   \quad  
	\Delta \mu_{\tau} = 4.2\%\ (3.5\%) \, ,  \quad
	\Delta \mu_{\mu}= 100\%\ (32\%)\,  		
\end{align}
at $\sqrt{s}=250$ GeV ($1$\,TeV) with $250$\,fb$^{-1}$ ($1$\,ab$^{-1}$).  
Also the TLEP presents a preliminary analysis~\cite{Gomez-Ceballos:2013zzn} of the expected 
precisions 
\begin{align}
	\Delta \mu_{b}= 0.2\% \, ,  \quad
	\Delta \mu_{c}=1.2\%,   \quad  
	\Delta \mu_{\tau} = 0.7\% \, ,  \quad
	\Delta \mu_{\mu}= 13\%\,  		
\end{align}
at $\sqrt{s}=240$\,GeV with $10$\,ab$^{-1}$ ($\Delta \mu_{c}$ is based on an 
extrapolation of the ILC~\cite{Baer:2013cma}).

Furthermore, there are ongoing discussions on whether it would be possible to also 
run precisely on the Higgs resonance and being able to measure the electron Yukawa~\cite{discussion1}.
The above information is based on the inclusive approach to particle identification. 
As one cannot apply $u,d,s$-jet-tagging with reasonable efficiencies, no direct 
information can be extracted on the Higgs coupling to these light-quark states.

%%%%%%%%%%%%%%%%%%%%%%%%%%%%%%%%%%%%%%%%%%%%%%%%
\section{Exclusive Higgs decays} \label{sec:Exclusive}
%%%%%%%%%%%%%%%%%%%%%%%%%%%%%%%%%%%%%%%%%%%%%%%%

Recently, the ATLAS collaboration provided the first upper bound on the rate 
for exclusive Higgs decays in the $h\to J/\psi\,\gamma$ mode, 
$\sigma_h\,\BR_{J/\psi\,\gamma}<33\,\fb\,$~\cite{Aad:2015sda}. 
This result is interesting not only because it can be interpreted as a bound 
on the Higgs couplings, in particular on the charm Yukawa~\cite{Perez:2015aoa}, 
but also because other exclusive decay modes are potentially subject to similar 
backgrounds.
In Ref.~\cite{Aad:2015sda} it is stated that the main background is inclusive 
quarkonium production where a jet in the event is reconstructed as a photon. 
This knowledge allows us to estimate the reach of future searches for exclusive 
Higgs decays in the $J/\psi\,\gamma$ channel as well as in other modes such as $\phi\gamma$
by using a {\tt PYTHIA} simulation of the backgrounds and rescaling to the ATLAS data.
We also note that we find a sizeable contribution to the background from 
a real photon and QCD $J/\psi$ or $\phi$ production in addition to the
jet-conversion background.

The Higgs signal-strength measurements at the LHC are sensitive only to the product 
of production cross section times branching ratio to a specific final state. 
The dependence on both the production and the total width cancels to good approximation 
in the ratio between the rates of two processes with similar production 
but different final states. 
In particular, we choose to normalize the exclusive decay signal strength 
by $h\to ZZ^*\to 4\ell$~\cite{Perez:2015aoa},
\begin{align}
	\label{eq:RMdef}
	\cR_{M\gamma,Z}
\equiv  \frac{{\mu}_{M\gamma}}{\mu_{ZZ^*}}\frac{\BR^{\rm SM}_{M\gamma} }{\BR^{\rm SM}_{ZZ^*\to 4\ell}}  
%=\frac{\sigma_h \BR_{M\gamma}}{\sigma_h \BR_{ZZ^*\to4\ell}}
\simeq \frac{ \Gamma_{M\gamma} }{ \Gamma_{ZZ^*\to 4\ell} }
= 	\left\{\begin{matrix}
		2.8 \times 10^{-2}\left( \kappa_\gamma - 8.7\times 10^{-2} \kappa_c \right)^2/\kappa_V^2  & \text{for}~M=J/\psi  \\
		2.4 \times 10^{-2}\left( \kappa_\gamma - 2.6\times 10^{-3} \kappa_s \right)^2/\kappa_V^2  & \text{for}~M=\phi 
	\end{matrix}\right.\, ,
\end{align}
where $\mu_{M\gamma}= {\sigma}_{h}{\BR}_{M\gamma}/\sigma_{h}^{\rm SM}\BR^{\rm SM}_{M\gamma}$
and $\kappa_X\equiv y_X/y^{\rm SM}_X$ and $V=Z,W$. Here, we assumed a perfect cancellation of the production cross sections and that the Higgs decay width to a $Z$ and two leptons ({\it e.g.} $h\to Z\gamma^*\to 4\ell$) is close to its SM value. The theoretical predictions for $h\to J/\psi\,\gamma$ and $h\to\phi\gamma$ are taken 
from Ref.~\cite{Bodwin:2014bpa} and \cite{Kagan:2014ila}, respectively, using the SM predictions  
$\BR^{\rm SM}_{J/\psi\,\gamma}=2.9\times10^{-6}$~\cite{Bodwin:2014bpa} and  
$\BR^{\rm SM}_{\phi\gamma}=3.0\times10^{-6}$~\cite{Kagan:2014ila}.
Ref.~\cite{Heinemeyer:2013tqa} gives $\BR^{\rm SM}_{ZZ^*\to4\ell}=1.25\times 10^{-4}$.

We are now in a position to study the prospects of the exclusive modes in the 
next phases of the LHC, HL-LHC, and a future $100$\,TeV $pp$ collider.
We first define the inequality,
\begin{align}
	\label{eq:Rbound}
	\cR_{M\gamma,Z} 
<	\frac{{\mu}^{95}_{M\gamma,E}}{\mu_{ZZ^*}} \frac{\BR^{\rm SM}_{M\gamma} }{\BR^{\rm SM}_{ZZ^*\to4\ell}} \, , 
\end{align}
where ${\mu}^{95}_{M\gamma,E}$ is a $95\%$\,CL upper bound for the $h\to M\gamma$ channel 
at the energy of $E=8,14,100$\,TeV. 
We neglect the uncertainty in $\mu_{ZZ^*}$, because it is expected to be smaller 
than $10\%$~\cite{CMS:2013xfa,ATLAS-collaboration:2012iza}. 
The inequality in Eq.~\eqref{eq:Rbound} together with Eq.~\eqref{eq:RMdef} leads to the 
following bound for the charm and strange Yukawa couplings,
\begin{align}
	\label{eq:kcrange}
	11\kappa_\gamma - 10 \kappa_V \left(\frac{\mu^{95}_{J/\psi \gamma,E}}{\mu_{ZZ^*}}\right)^{1/2} 
	< &\,\kappa_c <
	11\kappa_\gamma +10 \kappa_V \left(\frac{\mu^{95}_{J/\psi  \gamma,E}}{\mu_{ZZ^*}}\right)^{1/2}  \, ,  \\
	\label{eq:ksrange}
	3.8 \kappa_\gamma - 3.8 \kappa_V \left(\frac{\mu^{95}_{\phi  \gamma,E}}{\mu_{ZZ^*}}\right)^{1/2}  
	< &\, \frac{\kappa_s}{100} <
	3.8 \kappa_\gamma + 3.8 \kappa_V \left(\frac{\mu^{95}_{\phi  \gamma,E}}{\mu_{ZZ^*}}\right)^{1/2}  \, . 
\end{align}
If the upper bounds on the $J/\psi\,\gamma$ and $\phi\gamma$ signal strengths are
similar, the resulting bound on $\kappa_s$ is weaker than the bound on $\kappa_c$ by a factor 
of $\cO\left[ (m_c/m_s)\times  (m_{J/\psi}/m_\phi) \right]\,$.

We note in passing that during the last preparation stage of this paper, Ref.~\cite{Koenig:2015pha} appeared.
The authors of Ref.~\cite{Koenig:2015pha} presented an interpretation of the same ATLAS 
exclusive Higgs decay result, and obtained a weaker bound than the one found in Ref.~\cite{Perez:2015aoa}. 
The reason for this is threefold: (i) we normalised the signal strength of the exclusive 
channels by $\mu_{ZZ^*}$ to reduce the dependence on $\kappa_\gamma$ 
(which is more sensitive to new-physics contributions) while~\cite{Koenig:2015pha} chose 
to normalise it by $\mu_{\gamma\gamma}$, which leads to weakening the bound by $10\%$. 
This happens because the observed central value of $\mu_{\gamma\gamma}$~\cite{Aad:2014eha} is 
smaller than that of the ATLAS $\mu_{ZZ^*}$ result.  
(ii) we did not include the order $10\%$ theoretical uncertainty in the bound. 
(iii) most importantly, Ref.~\cite{Koenig:2015pha} has provided an improved and more 
precise calculation of the central value of the relevant matrix element that leads 
to a significant $40\%$ reduction in the dependence of $\kappa_c$ 
(and a slight increase of the theoretical uncertainties), which translates to a $40\%$ increase
in the bound.  

Next, we move to provide a rough estimation of the future bound on the 
$pp\to h \to J/\psi\,\gamma$ rate given the current ATLAS upper bound~\cite{Aad:2015sda}.  
We denote by $S^{95}_E$ the $95\%$\,CL upper bound on the number of signal events and 
by $B_E$ the expected number of background events at the center-of-mass energy $E$. 
Based on the available $8$\,TeV result ($S^{95}_8$), we estimate the future sensitivity by
assuming that
\begin{align}
	\frac{S_{E}^{95}}{\sqrt{B_{E}}}\approx \frac{S_{8}^{95}}{\sqrt{B_{8}}}\,. 
\end{align}
Using this, we find the following scaling,
\begin{align} \label{eq:BarMuS}
	{\mu}_{J/\psi \gamma,E}^{95} 
=	\frac{S_E^{95}}{S^{\rm SM}_{E}}
\approx 
	 \left(\frac{B_E}{B_8}\frac{S_8^{\rm SM}}{S_E^{\rm SM}}\right)^{\! 1/2}\!\!
 	\left(\frac{S_8^{\rm SM}}{S_E^{\rm SM}} \right)^{\!1/2}\!
	\frac{S_8^{95}}{S_8^{\rm SM}}
=
	\frac{1 }{R_{E}^{1/2} }\left(   \frac{\sigma^{\rm SM}_{h,8}~\cL_8}{\sigma^{\rm SM}_{h,E}~\cL_E}   \right)^{1/2}
	\mu^{95}_{J/\psi \gamma,8} \, ,
\end{align}
where
\begin{align}
	R_{E} \equiv \frac{S^{\rm SM}_E/B_E}{S^{\rm SM}_8\,/\,B_8}
	\qquad\text{and}\qquad
	\mu^{95}_{J/\psi\gamma,8}=\frac{S_8^{95}}{S_8^{\rm SM}} \,.\label{RE}
\end{align}
$S^{\rm SM}_{E}$ is the number of signal events as expected in the SM, 
$\sigma^{\rm SM}_{h,E}$ is the SM Higgs-production cross section 
and $\cL_{E}$ is the integrated luminosity. 
We have implicitly assumed above that the signal and background efficiencies are 
equal across the different runs. 
If future findings indicate that the efficiencies differ from each other, 
then the corresponding modification to Eq.~\eqref{RE}  can be absorbed by an 
appropriate rescaling of $R_{E}\,$.
The rate for Higgs production is characterised by a harder physical scale than 
the one of the corresponding QCD background. 
Consequently, colliders with larger center of mass are expected to have a 
larger signal to background ratio, i.e., $R_{E}\gtrsim1\,$.   

The expected upper bound on the signal strength in Eq.~\eqref{eq:BarMuS} can be 
easily interpreted as a bound on the Higgs couplings using Eq.~\eqref{eq:kcrange}. 
For $pp$ colliders with a center-of-mass energy of $14$\,TeV and $100$\,TeV, 
assuming $\mu_{ZZ^*}=\kappa_\gamma=\kappa_V=1$ and SM Higgs production, we find 
that the expected reach at $95$\%\,CL is 
\begin{align}
	\label{eq:kcexclusive14}
	11 - 80\left( \frac{1}{R_{14}}  \frac{ 2\times300\,{\rm fb}^{-1}}{\cL_{14}}\right)^{1/4} 
	<\,& \kappa_c < 
	11 + 80\left( \frac{1}{R_{14}} \frac{ 2\times300\,{\rm fb}^{-1}}{\cL_{14}}\right)^{1/4} \, , \\
	\label{eq:kcexclusive14-2}
	11 - 45\left( \frac{1}{R_{14}}  \frac{ 2\times3000\,{\rm fb}^{-1}}{\cL_{14}}\right)^{1/4} 
	<\,& \kappa_c < 
	11 + 45\left( \frac{1}{R_{14}} \frac{ 2\times3000\,{\rm fb}^{-1}}{\cL_{14}}\right)^{1/4} \, , \\[1em]
	\label{eq:kcexclusive100}
	11 - 40\left( \frac{1}{R_{100}}  \frac{ 2\times300\,{\rm fb}^{-1}}{\cL_{100}}\right)^{1/4} 
	<\,& \kappa_c < 
	11 + 40\left( \frac{1}{R_{100}}  \frac{ 2\times300\,{\rm fb}^{-1}}{\cL_{100}}\right)^{1/4} \, ,\\
	\label{eq:kcexclusive100-2}
	11 - 22\left( \frac{1}{R_{100}}  \frac{ 2\times3000\,{\rm fb}^{-1}}{\cL_{100}}\right)^{1/4} 
	<\,& \kappa_c < 
	11 + 22\left( \frac{1}{R_{100}}  \frac{ 2\times3000\,{\rm fb}^{-1}}{\cL_{100}}\right)^{1/4} \, .
\end{align}
Here, we used $ \sigma^{\rm SM}_{h,(8,14,100)}=22.3 ,\, 57.2,\, 897\,$pb~\cite{Heinemeyer:2013tqa}, 
$\cL_8=19.2\,\fb^{-1}$ and $\mu^{95}_{J/\psi,8}=515$~\cite{Aad:2015sda}. 
These bounds may be compared to the current bound of $\kappa_c\lesssim 220\,$~\cite{Perez:2015aoa}. 
We see that the projected bounds depend only weakly on the integrated luminosity and on $R_{E}$. 
The corresponding expected upper bound on the branching ratio is 
$\BR_{J/\psi, 14(100)}<2.4\,(0.60)\times 10^{-4}$, where we assume SM production and 
$\cL_{14(100)}=300\,\fb^{-1}$.  

The different exclusive channels are expected to be subject to analogous backgrounds, namely 
QCD production and an associated fake jet or a real photon.  
The ATLAS result for $h\to J/\psi \gamma$~\cite{Aad:2015sda} can be, thus, used to estimate 
the future reach in the different channels. 
In particular, we focus on the case of  $h\to\phi\gamma\,$ decay, but the generalisation of 
our analysis to other final states, such as $\rho$ or $\omega$ is straightforward.
However, as our results are very pessimistic we do not expect good results for the other
analyses.

To make the following discussion more transparent, we supplement our previously used symbols 
for signal and background ($S$, $B$) with a subscript $J/\psi\gamma$ ($S_{J/\psi, E}$ and 
$B_{J/\psi,E}$). Symbols regarding $h\to \phi \gamma$  will contain a $\phi\gamma$ subscript.

In order to estimate the upper bound on the $h\to\phi\gamma$ signal strength, 
we use an approximation, 
\begin{align}\label{eq:phiS95}
	\frac{S^{95}_{\phi\gamma,E}}{\sqrt{B_{\phi\gamma, E}} }  \approx  \frac{S^{95}_{J/\psi\gamma, E}}{\sqrt{B_{J/\psi\gamma, E}} } \ .
\end{align}
We then estimate $S_{\phi\gamma, E}$ and $B_{\phi\gamma, E}$ in the following way.
The ratio between the number of signal events in each channel is given by
\begin{align} \label{eq:SphiJpsi}
	\frac{S_{\phi\gamma, E}}{S_{J/\psi\gamma, E}} 
=&	\frac{ \sigma_{h,E}\, \BR(h\to \phi\gamma) \,\cL_E }
	{\sigma_{h,E}\, \BR(h\to J/\psi\,\gamma) \,\cL_E }
	\frac{\BR(\phi\to K^+ K^-)}{\BR(J/\psi\to \mu^+\mu^-)}\frac{\epsilon_\phi}{\epsilon_{J/\psi}}
	\, 
\end{align}
where $\epsilon_{J/\psi(\phi)}$ stands for the triggering and reconstruction efficiency 
(including the isolation and various kinematical cuts following Ref.~\cite{Aad:2015sda}). 
The $J/\psi$ is observed via its rather clean $J/\psi \to \mu^+\mu^-$ leptonic decay mode, while 
the $\phi$ is assumed to decay to $\phi\to K^+K^-$, which is a much more challenging final state 
for triggering, identification and background rejection.
Nevertheless, we focus on this final state because it has a large branching ratio.
In that sense, the bound below is rather conservative, given that we ignore these 
challenges when we rescale the ATLAS $J/\psi$ result.
The ratio of backgrounds for the two different exclusive final states can be written as
\begin{align} \label{eq:BphiJpsi}
	\frac{B_{\phi\gamma, E}}{B_{J/\psi, E}}
=	\frac{ \sigma_E(pp\to\phi\,``\gamma")  }
	{\sigma_E(pp\to J/\psi\,``\gamma")  }
	\frac{\BR(\phi\to K^+ K^-)}{\BR(J/\psi\to \mu^+\mu^-)}\frac{\epsilon_\phi}{\epsilon_{J/\psi}}
	\,,
\end{align}
where by $``\gamma"$ we refer to a photon candidate, namely an object that has passed the 
(ATLAS) tight-photon selection cuts, i.e., it is either a genuine photon or a jet 
faking a photon. The corresponding rate reads 
$$\sigma_E(pp\to\phi,J/\psi\,``\gamma")=\sigma_E(pp\to\phi,J/\psi\,j)  \,P(j\to\gamma) +\sigma_E(pp\to\phi,J/\psi\,\gamma)\,, $$  
where $P(j\to\gamma)\sim2\times10^{-4}$ stands for the rate that a jet is 
misidentified as a photon under the tight photon selection~\cite{ATLAS:2011kuc}. 

Combining Eq.~\eqref{eq:phiS95} with Eqs.~\eqref{eq:SphiJpsi}--\eqref{eq:BphiJpsi} leads 
to an extrapolation of the upper bound for the $h\to\phi\gamma$ signal strength,
\begin{align}
	\label{eq:muphi}
	\mu^{95}_{\phi\gamma, E}  
	=\frac{S_{\phi\gamma, E}^{95}}{S_{\phi\gamma, E}^{\rm SM}}
	&\approx	\nonumber
	\left(\frac{B_{\phi\gamma,E}}{B_{J\gamma,E}}\right)^{\!1/2}\frac{S_{J\psi\gamma, E}^{\rm SM}}{S_{\phi\gamma, E}^{\rm SM}}
	\frac{S_{J\psi\gamma, E}^{95}}{S_{J\psi\gamma, E}^{\rm SM}}
\\&=	\left(\frac{ \sigma_E(pp\to \phi\,``\gamma")}{\sigma_E(pp\to J/\psi\,``\gamma")}
	\frac{ \BR(J/\psi\to \mu^+\mu^-)}{\BR(\phi\to K^+ K^-) }
	\frac{ \epsilon_{J/\psi} }{\epsilon_{\phi} }  \right)^{\!1/2}		
	\frac{\BR^{\rm SM}_{J/\psi\gamma}}{ \BR^{\rm SM}_{\phi\gamma} }\
	\mu^{95}_{J/\psi\gamma, E}  \nonumber
	\\&=
	0.34 \,  \left(\frac{ \sigma_E(pp\to \phi\,``\gamma")}{\sigma_E(pp\to J/\psi\,``\gamma")}
	\frac{\epsilon_{J/\psi}}{\epsilon_{\phi}}\right)^{1/2}	\mu^{95}_{J/\psi,E}
\end{align}
where $\BR(\phi\to K^+ K^-)=48.9\%$ and $\BR(J/\psi\to \mu^+\mu^-)=5.93\%$~\cite{PDG}. 
As the $J/\psi$ is reconstructed from a dimuon pair while the $\phi$ from a $K^+K^-$ pair, 
we expect that $\epsilon_{J/\psi}>\epsilon_\phi\,$. Moreover, we expect that 
$\sigma_E(pp\to \phi\,j)>\sigma_E(pp\to J/\psi\,j)$ because $J/\psi$'s are more rarely 
produced than
$\phi$ in the QCD process. 
Therefore, we expect the upper bound for $h\to \phi\gamma$ to be weaker than that 
for $h\to J/\psi\gamma$, more precisely $\mu^{95}_\phi\gg0.34\,\mu^{95}_{J/\psi}\,$. 

The estimation of the ratio of $\sigma_E(pp\to \phi\, ``\gamma")/\sigma_E(pp\to J/\psi\,``\gamma")$ 
was performed in two steps.
Using {\tt PYTHIA~8.2}~\cite{Sjostrand:2006za,Sjostrand:2014zea} we generated two samples. 
The first of events with a photon and a jet and the second one with di-jets at $\sqrt{s}=8\,$TeV. 
Prior to showering and hadronization, we required that the two objects have $p_T>20\,$GeV 
at the parton level.
Following Ref.~\cite{Aad:2015sda}, we then selected events that contain $J/\psi (\phi)$ with 
$p_T>36\,$GeV, a photon or an anti-$k_T$ jet (of cone $0.4$) with $p_T>36\,$GeV, $\abs{\eta}<2.37$ 
and $\Delta \phi(J/\psi(\phi), ``\gamma")>0.5\,$, where $``\gamma"$ stands either for the second 
jet multiplied by $P(j\to\gamma)\sim 2\times10^{-4}$ \cite{ATLAS:2011kuc} or the actual photon. 
Following the ATLAS analysis, we also required an isolation cut for the $J/\psi$ ($\phi$). 
We evaluated the sum of the energy of the extra hadrons (or photons from $\pi^0$ decay) that 
are within a cone of $0.2$ away form the  $J/\psi$ ($\phi$) and required it to be less than 
$10\%$ of the energy of the $J/\psi$ ($\phi$).
We found that the number of events that pass these cuts for the fake jet sample is very close 
to that of the real photon, and thus retained both samples.
The resulting ratio from our simulation is 
\begin{align}
	\label{eq:RsigmaPythia}
	\left. \frac{ \sigma_8(pp\to \phi\,``\gamma")}{\sigma_8(pp\to J/\psi\,``\gamma")} \right|_{\tt PYTHIA} \sim \, 9.
\end{align}
We have verified that for the $jj$ sample a similar ratio has been obtained for the $\sqrt{s}=14\,$TeV case. As a sanity check we compare our total simulated rate of $\sigma(pp\to J/\psi\,``\gamma")$ with the one reported by ATLAS and to agree within $\sim 50\%$. 

All the above information can be combined to constrain $\kappa_s$. 
Using Eqs.~\eqref{eq:ksrange}, \eqref{eq:muphi} and~\eqref{eq:RsigmaPythia} we find,
\begin{align}
	3.8  - 29\left( \frac{1}{R_{14}}  \frac{ 2\times300\,{\rm fb}^{-1}}{\cL_{14}}\right)^{1/4} 
	<\,& \frac{\kappa_s}{100} < 
	3.8  + 29 \left( \frac{1}{R_{14}}  \frac{ 2\times300\,{\rm fb}^{-1}}{\cL_{14}}\right)^{1/4} \, , \\
	3.8 - 16\left( \frac{1}{R_{14}}  \frac{ 2\times3000\,{\rm fb}^{-1}}{\cL_{14}}\right)^{1/4} 
	<\,& \frac{\kappa_s}{100} < 
	3.8 + 16\left( \frac{1}{R_{14}}  \frac{ 2\times3000\,{\rm fb}^{-1}}{\cL_{14}}\right)^{1/4} \, , \\
	3.8 - 14\left( \frac{1}{R_{100}}  \frac{ 2\times300\,{\rm fb}^{-1}}{\cL_{100}}\right)^{1/4} 
	<\,& \frac{\kappa_s}{100} < 
	3.8 + 14\left( \frac{1}{R_{100}}  \frac{ 2\times300\,{\rm fb}^{-1}}{\cL_{100}}\right)^{1/4} \, , \\
	3.8 - 8.2\left( \frac{1}{R_{100}}  \frac{ 2\times3000\,{\rm fb}^{-1}}{\cL_{100}}\right)^{1/4} 
	<\,& \frac{\kappa_s}{100} < 
	3.8 + 8.2\left( \frac{1}{R_{100}}  \frac{ 2\times3000\,{\rm fb}^{-1}}{\cL_{100}}\right)^{1/4} \, ,
\end{align}
where we assumed $\mu_{ZZ^*}=\kappa_\gamma=\kappa_V=1\,$. Since the only known possibility to probe the strange Yukawa in hadron machines is via $h\to\phi\gamma$, the resulting reach is still weak, $\cO(\kappa_s)\lesssim2000$ even 
at HL-LHC. However, this disappointing situation may be improved by the development of new methods 
that would take advantage of the relatively quiet QCD environment of a Higgs event. 
For instance, one direction is to consider jet-substructure techniques.

The first-generation Yukawas may be probed via the $h\to(\rho ,\, \omega)\,\gamma$ 
decays~\cite{Kagan:2014ila}. 
However, since the $\rho$ and $\omega$ mesons are lighter than $\phi$, we expect  
a larger QCD background in hadron colliders, resulting in weaker sensitivity. 
Nevertheless, due to the large $\gamma-\rho$ mixing,  the process $h\to\rho\gamma$ 
has a branching ratio, $\BR^{\rm SM}_{\rho\gamma}=1.9\times10^{-5}$, larger than in 
the other modes, and may be probed in the clean environment of future $e^+e^-$ colliders. 

It is interesting to compare the projected sensitivity reach on $\kappa_c$ 
between the inclusive rate with $c$-tagging and the exclusive decays to $h\to J/\psi\,\gamma$. 
From Tables~\ref{tab:profiles14TeV}, \ref{tab:profiles100TeVboosted}, \ref{tab:profiles100TeVunboosted} 
and Eqs.~\eqref{eq:kcexclusive14}, \eqref{eq:kcexclusive14-2}, \eqref{eq:kcexclusive100}, 
\eqref{eq:kcexclusive100-2}, we learn that the prospects of the inclusive analysis to probe the 
charm Yukawa are much better than of the exclusive analysis.  
For example, with  $300\,\fb^{-1}$ the projected reach in the inclusive analysis is 
stronger than in the exclusive analysis by roughly a factor of $4$, and  with $3000\,\fb^{-1}$ 
in the high-luminosity stage it is stronger by a factor of $10$. So given the current background understanding for $h\to J/\psi\,\gamma$, we expect 
that the inclusive $c$-tagging method will be more powerful in probing modifications of
the charm Yukawa.\\

%%%%%%%%%%%%%%%%%%%%%%%%%%%%%%%%%%%%%%%%%%%%%%%%
\section{Conclusions} \label{sec:conc}
%%%%%%%%%%%%%%%%%%%%%%%%%%%%%%%%%%%%%%%%%%%%%%%%

In this work, we have presented the projections for probing light-quark Yukawas within the LHC, 
its high-luminosity stage~(HL-LHC) and a future $100$\,TeV hadron collider. 
Using charm tagging we find that the HL-LHC can probe the charm Yukawa with a
sensitivity of a few times the SM value. With an improved tagger at a $100$\,TeV machine 
a sensitivity close to the one necessary for probing the SM value appears feasible.

We have also provided a preliminary study of the sensitivity of the various exclusive decay modes.
These channels are particularly important as they provide a unique opportunity to probe the Higgs 
couplings to the three lightest quarks. 
Our study shows that the reach of the exclusive modes, however, is rather limited,
as follows.
ATLAS recently provided the first measurement of the background relevant to charmonia 
and a photon final state; a final state that can potentially also probe the 
Higgs--charm coupling. 
ATLAS observes a large continuous background due to QCD production of charmonia and a jet 
converted into a photon. 
We also find, using a leading-order simulation, a sizeable contribution from  charmonia plus 
a photon production. 
Given the small signal and the large background we find that the reach based on the 
$J/\psi\, \gamma$ mode is more than an order of magnitude weaker than that
expected in the inclusive approach. 

Focusing on analogous backgrounds, we further study the sensitivity reach of 
the search for the $\phi \gamma$ final state, which can probe the Higgs to strange coupling.
In this case, the resulting sensitivity is poorer, allowing the HL-LHC to only prove a strange 
Yukawa of order $10^3$ times the SM value or more. The current analysis strategy for the exclusive modes is subject to large backgrounds. These consist of pure QCD production (with a jet faking a photon) 
and QCD plus associated-photon production that limit the reach of the analysis.
However, these backgrounds are not irreducible. 
The situation may be improved as follows. One can modify the search, limiting 
the kinematics such that the Higgs is boosted and captured by a fat jet. 
In this case, the energy deposition for the signal and background inside the fat jet 
should be very different leading to a better background rejection and an improved 
sensitivity to the Yukawa couplings. 

%%%%%%%%%%%%%%%%%%%%%%%%%%%%%%%%%%%%%%%%%%%%%%%%
\section*{Acknowledgments} 
%%%%%%%%%%%%%%%%%%%%%%%%%%%%%%%%%%%%%%%%%%%%%%%% 
We thank David Kosower for useful discussions. We  acknowledge help from the ATLAS 
collaboration for providing us with details of the analysis of Ref.~\cite{ATL-PHYS-PUB-2014-011}.
The work of KT is supported in part by the Grant-in-Aid for JSPS Fellows, 
the work of GP is supported by ERC, IRG and ISF.  

%%%%%%%%%%%%%%%%%%%%%%%%%%%%%%%%%%%%%%%%%%%%%%%%
%%%%%%%%%%%%%%%%%%%%%%%%%%%%%%%%%%%%%%%%%%%%%%%%
\appendix
%%%%%%%%%%%%%%%%%%%%%%%%%%%%%%%%%%%%%%%%%%%%%%%%
%%%%%%%%%%%%%%%%%%%%%%%%%%%%%%%%%%%%%%%%%%%%%%%%

%%%%%%%%%%%%%%%%%%%%%%%%%%%%%%%%%%%%%%%%%%%%%%%%
\section{Supplement for inclusive analysis} \label{app:inclusive}
%%%%%%%%%%%%%%%%%%%%%%%%%%%%%%%%%%%%%%%%%%%%%%%%
\subsection{Correlation between $b$- and $c$-tagging}
\begin{figure}[t!]
\centering
\vspace{-10pt}
\includegraphics[]{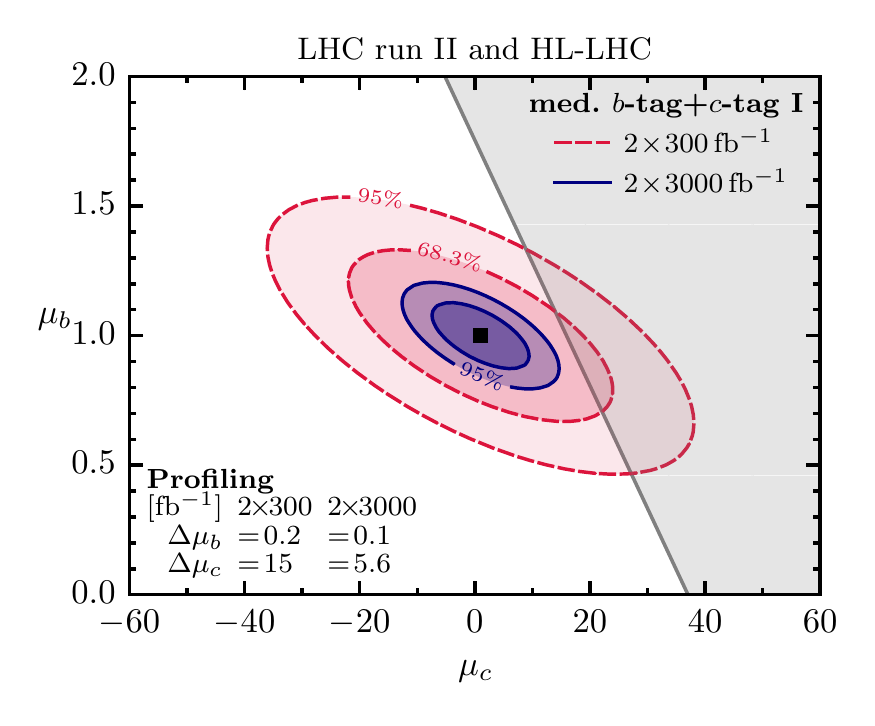}
\hspace*{-1em}
\includegraphics[]{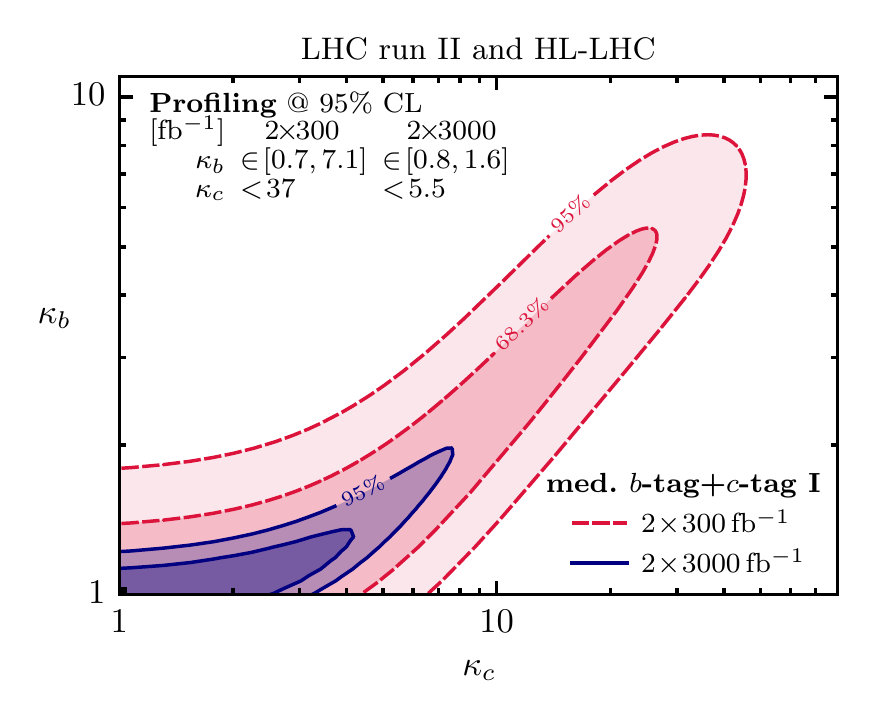}
\vspace{-10pt}
\caption{$300$\,fb$^{-1}$ and $3000$\,fb$^{-1}$ prospects for the signal strengths and couplings of 
$h\to b\bar b$ and $h\to c\bar c$ at the LHC based on $b$- and $c$-tagging I for the correlated scenario.  
 Future sensitivity of $\mu_b$ and $\mu_c$ is in the left panel. The grey shaded region is unphysical unless Higgs production is modified with respect to the SM case. Future sensitivity $\kappa_b$ and $\kappa_c$ is in the right panel. All other Higgs couplings are assumed to be like in the SM. 
The profiled likelihood ratio \cite{Cowan:2010js} is used for the respective reach of $\Delta\mu_b$, $\Delta\mu_c$, $\kappa_b$ and $\kappa_c$.}
\label{fig:mub-muc-corr}
\end{figure}

When $b$- and $c$-taggers are simultaneously imposed, we study two scenarios with 
respect to correlations.  The actual situation is expected to be something between the two scenarios. 
In the main text, we showed only the uncorrelated scenarios, and here we show in
Fig.~\ref{fig:mub-muc-corr} the projection using $b$-tagging and $c$-tagging I in the correlated scenario.
Comparing to  Fig.~\ref{fig:mub-muc} (left) and Fig.~\ref{fig:kab-kac} (left), we find 
the results in the two scenarios to be very similar. This is for the following reason. 
The main difference of the two scenarios is whether the two $b$-tagged category (i) contains  
$c$-tagged events or not. 
However, this statistical difference does not change the overall number in category (i), 
and hence it does not affect the main significance of the category (i) that determines $\mu_b$. 
In addition, the two $c$-tagged category (iii), which is the most important to measure 
$\mu_c$ is exactly the same in both scenarios. Therefore, the two scenarios give almost 
the same sensitivity.

\begin{figure}[t!]
\centering
%\vspace{-10pt}
\includegraphics[width=0.3\linewidth]{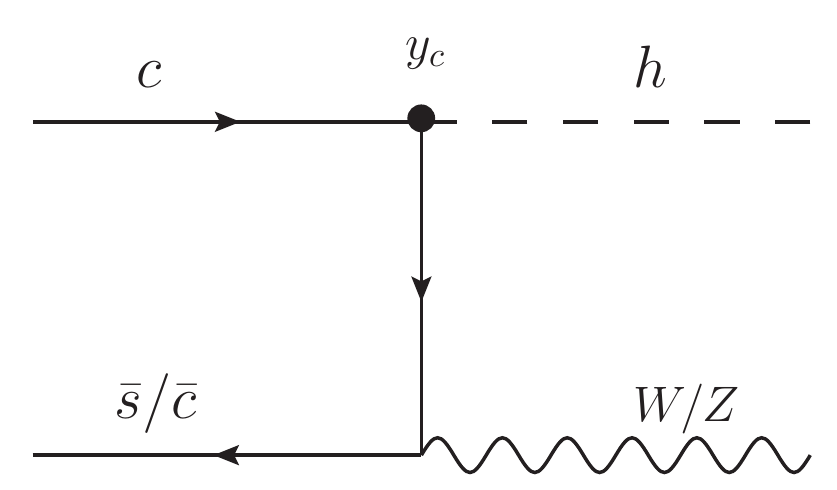}
%\vspace{-10pt}
\caption{Example diagram that modifies $Vh$ production when the charm-quark Yukawa 
is enhanced.
\label{fig:Vhdiag}}
\end{figure}
%

%%%%%%%%%
 \begin{figure}[t!]
\centering
%\vspace{-10pt}
\includegraphics[width=0.45\linewidth]{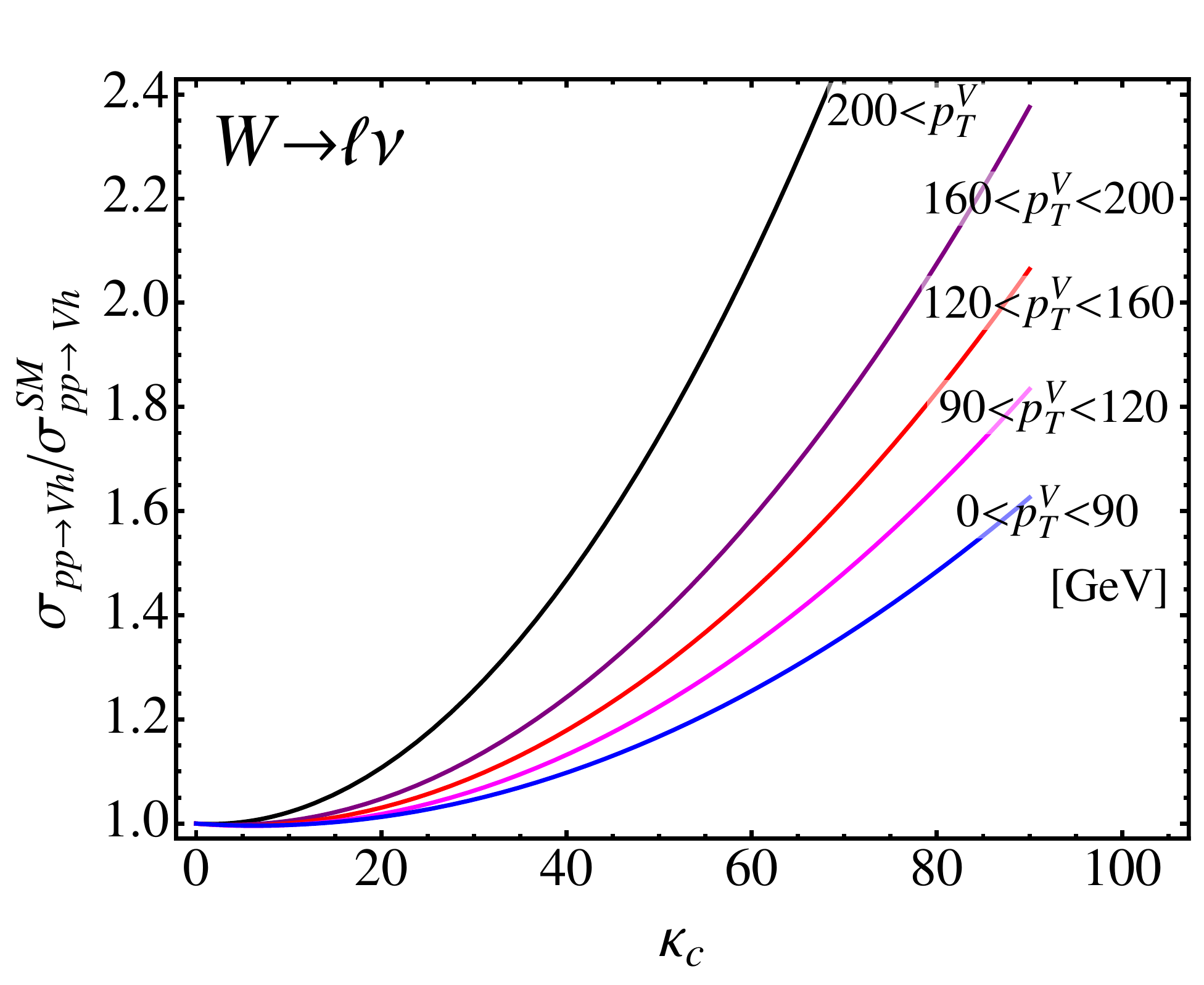}
\hspace*{1em}
\includegraphics[width=0.45\linewidth]{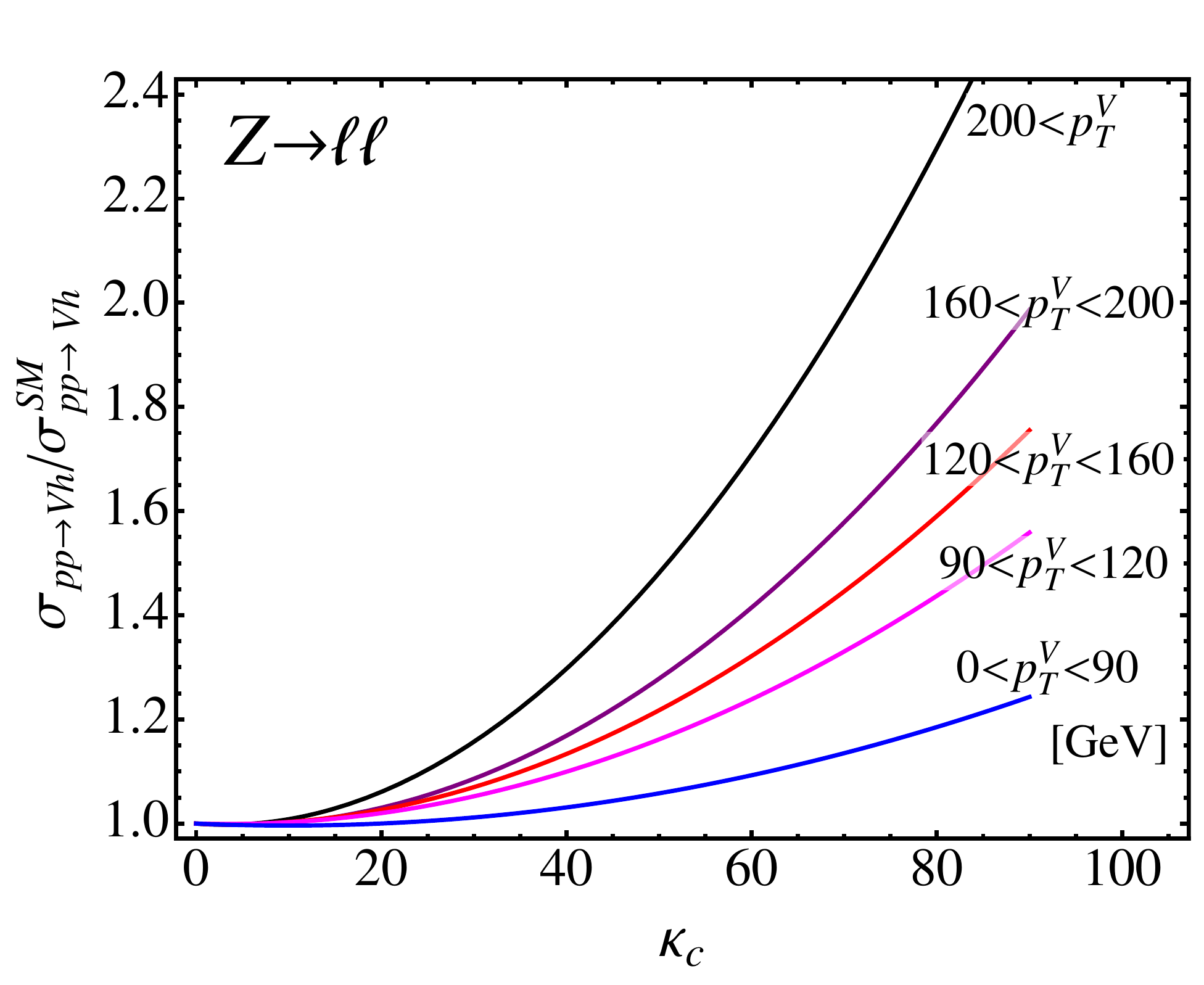}
\caption{$Vh$ enhancement with $\kappa_c$ from the new production mechanism, using the selection cuts  ATLAS.
\label{fig:VhNonSM}}
\end{figure}
%%%%%%%%%

For $c$-tagging II and III, we can only assume the uncorrelated scenario 
because the assumption for the correlated scenario leads to an inequality, 
\begin{equation}
\epsilon_{x}^{\text{($c$-tag)}}  \leq \epsilon_{x}^{\text{($b$-tag)}} \ \  \text{for all } x=b,c,l.   
\end{equation}
$c$-tagging II or III with medium $b$-tagging does not satisfy this inequality. 
However, the $c$-tagger is expected to be less correlated with $b$-tagging after 
its possible improvement, so it is reasonable to expect the uncorrelated scenario 
to be realistic. 

We emphasize that the correlation of the two taggers are not an issue for the actual
experiments because they have the information for the respective jet. 
We studied the two different scenarios simply because the information is not available 
for us to take into account the correlation.   

%%%%%%%%%%%%%%%%%%%%%%%%%%%%%%%%%%%%%%%%%%%%%%%%
\subsection{Non-Standard Model $Vh$ production}% \label{app:NonSMVh}
%%%%%%%%%%%%%%%%%%%%%%%%%%%%%%%%%%%%%%%%%%%%%%%%
For large $\kappa_c \sim\!\cO(10{\rm -}100)$, new contributions to the $Vh$ final 
states, shown in Fig.~\ref{fig:Vhdiag}, become important and the signal strength 
is modified not only by the branching ratio but also by the production cross section. 
The contributions to the $Vh$ production cross section at $\sqrt{s}=14$\,TeV as a function of 
$\kappa_c$ are presented in Fig.~\ref{fig:VhNonSM} and are roughly given by 
\begin{equation}
	\frac{\sigma_{pp\to Vh}}{\sigma^{\rm SM}_{pp\to Vh}}\simeq1 +\left(\frac{\kappa_c}{60{\rm -}180}\right)^2\ 
\end{equation}
for large $\kappa_c$. Here, the Higgs coupling to the $W/Z$ is assumed to be SM like, i.e. $\kappa_V=1$. 
We obtained these results using {\tt MadGraph} 5.2~\cite{Alwall:2011uj} at the parton level and 
leading order, applying the ATLAS selection cuts for the LHC $14\,$TeV run~\cite{ATL-PHYS-PUB-2014-011}. 

For 100\, TeV $pp$ collider, the effect of the new production mechanism is not 
important because the sensitivity in $\kappa_c$ is $\cO(1)$. Therefore, we neglect this effect.

%%%%%%%%%%%%%%%%%%%%%%%%%%%%%%%%%%%%%%%%%%%%%%%%
%%%%%%%%%%%%%%%%%%%%%%%%%%%%%%%%%%%%%%%%%%%%%%%%
%\bibliographystyle{apsrev}
\bibliographystyle{aipnum4-1}
\bibliographystyle{apsrev4-1}
\bibliography{ref}
%%%%%%%%%%%%%%%%%%%%%%%%%%%%%%%%%%%%%%%%%%%%%%%%
%%%%%%%%%%%%%%%%%%%%%%%%%%%%%%%%%%%%%%%%%%%%%%%%

\end{document}